\documentclass{article}

\usepackage[title]{appendix}
\usepackage{setspace}
\usepackage{shortcuts}
\usepackage{PRIMEarxiv}

\usepackage[utf8]{inputenc} % allow utf-8 input
\usepackage[T1]{fontenc}    % use 8-bit T1 fonts
\usepackage{booktabs}       % professional-quality tables
\usepackage{amsfonts}       % blackboard math symbols
\usepackage{nicefrac}       % compact symbols for 1/2, etc.
\usepackage{microtype}      % microtypography
\usepackage{lipsum}
\usepackage{fancyhdr}       % header
\usepackage{graphicx}       % graphics
\graphicspath{{media/}}     % organize your images and other figures under media/ folder
\PassOptionsToPackage{hyphens}{url}\usepackage{hyperref}
\usepackage{adjustbox}
\usepackage{booktabs, threeparttable, dcolumn, multirow}

\usepackage{moresize}
\usepackage{etoolbox}
\AtBeginEnvironment{quote}{\par\singlespacing\small}
\usepackage{sankey}
\usepackage{rotating}
\usepackage{newfloat}
\usepackage{amsmath} 
\usepackage{algorithm}
\usepackage{algpseudocode}
\usepackage{natbib}
\usepackage{hyperref}
\hypersetup{
	colorlinks,
	linkcolor=black,
	citecolor=black,
	urlcolor=black
}

\usepackage{amssymb}
\title{How candidates evoke identity and issues on TikTok
}

\author{
  Sabina Tomkins \\
  University of Michigan \\
  \texttt{stomkins@umich.edu} \\
   \And
  Chang Ge \\
  University of Michigan \\
  \texttt{changgge@umich.edu} \\
   \And
  David Rothschild \\
  Microsoft Research \\
  \texttt{david@researchdmr.com } \\
}

\begin{document}
\maketitle

\begin{abstract}
Social media platforms are increasingly becoming a key aspect of campaign communications strategy, with both paid (advertising) and earned (organic) posts, working towards fundraising (of core supporters), get-out-the-vote (of supporters), and persuasion of (of non-supporters). TikTok, and other short-form video content platforms, with both a unique format (short-vertical videos) and algorithms (powered more by individual content than channel subscriptions) demand unique content creation. We explore the final six months before the 2024 US Presidential Election to understand the nature of what the major campaigns were producing.
We frame our analysis around two theories in political science. 

The first, is the \textbf{expressive (identities)} model of politics. Under this model, voters are motivated by the interests and standing of the groups they belong to and candidates would appeal to the social identities that are important to voters.  Alternatively, candidates may follow an \textbf{instrumental (issues)} model, which states that voters align with politicians who advocate for the issues which are important to them. 
For each of these models, and their combination, we also inspect the extent to which each candidate attacks the other. 
This speaks to literature around campaign marketing that broadly finds that attacks are common and meaningful for politics.

The next logical set of questions is how this messaging in TikTok aligns with the opinions of potential voters. That is, did the candidates speak about the identities and issues which describe the potential voters within their party? Alternatively, instead of speaking to the identities and issues which describe their potential voters, they may try to differentiate themselves 
from the other candidate by speaking to the identities/issues which are particularly important to voters in their party and \textbf{not} in the other.  

To understand how candidates positioned themselves on TikTok and how these positions represented and aligned with how people identify with their political party, we combine two unique datasets. 
The first is a collection of posts from the Harris and Trump campaigns on TikTok from July to November 2024. 
The second is the responses to a two-wave survey conducted in 2022 with $\sim$1,000 respondents. 
Combining these two datasets, we find that Harris and Trump employed different strategies on TikTok. 
Trump was more likely to disparage Harris and to speak to identities and issues which differentiate Republicans from Democrats, while Harris was more likely to speak to identities which represent Democratic voters and the issues which they value. Whereas we find that issues are more predictive of party ID and may thus make sense to speak about in campaign posts on TikTok, both candidates were more likely to mention identities (34\%  of posts on average) than issues (25\% of posts on average), and most likely to mention neither (55\% of posts on average).  
\end{abstract}

% keywords can be removed
\keywords{\textit{political identity, social groups, policy issues, partisanship, TikTok, campaign messaging, prediction} \vspace{8ex}}

\section{Introduction}
Adults, especially young adults, are increasingly encountering political information on social media. In traditional media environments, news is very siloed, where consumers actively choose to go to a news website or turn on a news channel.
People also choose not to consume news from traditional environments at all, whereas
on social media, news content can reach people who both choose to consume but also incidentally to allow the news to find them \citep{gil2017effects,enli2017twitter}. Social media is broadly recognized as a critical factor in elections \citep{eshbaugh2015traditional, rossini2017disentangling, bode2020words, groshek2013public, metzgar2009social, enli2017twitter, buccoliero2020twitter, fulgoni2016power}. Pew's recent TikTok study noted that 95\% of users are on the social media platform for entertaining, but 45\% see at least some political content \citep{pew2024}. Political content on social media thus has the potential to reach more marginal voters in terms of both persuasion and likeliness to vote, as well as to provide enhanced targeting for fundraising. TikTok, consistently one of the top mobile apps in the US, is thus a platform where users encounter political campaigns with implications for political behavior \citep{hindarto2022tiktok}. 

One important question about political activity on this platform is how candidates appeal to voters.
We argue that candidate behavior on this platform can be inspected through two political theories: the expressive (identities) and instrumental (issues) models of politics. Each model seeks to explain political behavior. However, to the best of our knowledge, there is no existing work on understanding which model candidates employ, and to what extent, on social media. This question has critical implications for political science, as these models provide fundamentally different ideas of politics, with the first arguing that voters choose the candidate that they most identify with as a person or which best expresses their social/political identity \citep{huddy2017political}, 
%achen2017democracy,huddy2016political
and the second arguing that voters choose candidates systematically according to agreement on issues \citep{fiorina1976voting,lenz2012follow,highton2011long}\footnote{Note, that the presence of these two theories does not prevent the existence of complementary theories which combine elements from each, nor does our analysis. We investigate each for the sake of understanding, while allowing that both potential voters and elites consider both identities and issues in their behavior.}.

These two themes have been used to understand political polarization. 
There is widespread concern around the phenomena of ``political tribalism'' and affective polarization \citep{campbell2018polarized,lelkes2016mass,iyengar2015fear} where positive affect for the out-party 
in the United States has steadily decreased in the last thirty years \citep{iyengar2019origins}.
While party identity is often constructed as a social identity \citep{mason2018one}, others have argued that policy agreement with a party's platform is a stronger driver of party attachment and affective polarization \citep{orr2023affective}. 
These two views have different implications for real-world behavior \citep{iyengar2015fear,shafranek2020political,huddy2015expressive}. 
%The strength of views have different potential implications for behavior outside of the political sphere, with some arguing that 
%the extent to which people identify socially with a party can have life or death consequences
Furthermore, politicians can potentially exasperate polarization with the messages they deploy on social media \citep{martin2024deepening}.
Candidates' posts on social media thus offer a glimpse into whether they believe identities or issues to be more effective frames, or if they tend to elicit both, with implications for both political and non-political behavior.

If people primarily prefer candidates because they represent the social groups they belong to, that implies that reasonable candidates should spend their time appealing to these identities by mentioning particular social groups. It also implies that popular policies will be those which reward and punish different groups. If, on the other hand, issues play a greater role in political judgment, politicians should speak about issues that matter to people, and focus their political energy on policies which act on the issues important to their constituents. 
Yet, there has been no systematic review of how \textit{candidates} position themselves according to identities and issues on social media. 
This work thus speaks to an open question of what partisan identity really is \citep{ruckelshaus2022kind}, through 
understanding elite behavior on social media and how that aligns with the attitudes of potential voters.

In describing candidate positioning on social media, we add one additional dimension: the types of attack content contained in a post. Such content has been extensively studied in the literature \citep{fridkin2004negative,lau1999effects,fine2023negativity,auter2016negative,de2012social,yu2024partisanship}, 
with some finding that negative messaging is effective at influencing behavior on social media \citep{fine2023negativity}.
However, while others have described the extent to which candidates attack the other relative to mentioning issues, no one has related the presence of attacks to the mentions of identities and issues. This is a striking gap, as considerable attention has been paid to the rise of identity politics \citep{fukuyama2018against}.

Finally, we ground our analysis of candidate behavior through the opinions of potential voters. 
That is, does the model which best describes candidates behavior also best describe self-reported party identification (party ID)? 
Do candidates speak to the identities and issues which are most common and important to potential voters within each party, 
or those which best differentiate them from the other party?
To answer these questions, we utilize survey data which both measures different aspects of social identity and participants' positions on different issues.
Thus, our focus is on the following research questions:

\begin{itemize}
    \item Research Question 1: Did the presidential candidates in the United States 2024 election speak to identities or issues in their TikTok posts? Which did they mention more? And, what was the interaction between identities and issues, and negative affect?
    \item Research Question 2: Which identities and issues did they discuss?
    \item Research Question 3: How do the identities and issues discussed by candidates align with the attitudes of potential voters within their political party?
   % \item Which best predicts party ID, identities or issues, and does the most predictive theory align with what is most discussed by candidates?  
\end{itemize}

To answer these questions, we combine a study of TikTok posts in the lead up to the 2024 election with responses to a two-wave survey completed in 2022. The first wave of the survey concentrated on political identities and the second on political issues. Combining these datasets allows us to understand not only what candidates did, but also how these actions correspond to the opinions and preferences of a representative sample of the American public (potential voters).

\section{Related Work}
Our work connects to several directions of study. We are primarily studying candidate messaging on social media, and we begin with other work that has studied political candidates and campaigns in online settings. 
Next, as our goal is to understand the extent to which these messages seem to imply that elites hold an expressive and/or instrumental view of politics, we describe these two theories. 
The secondary level of our taxonomy is concerned with the types of attacks candidates make, and thus we describe related work around negative advertising. 
Finally, we discuss what candidates may be trying to achieve on TikTok. 

\subsection{Candidate positioning and social media}
Campaign advertising is an essential mechanism for candidates to inform and mobilize voters \citep{freedman2004campaign}, 
where
existing work has found that 
candidates' branding can influence voting decisions in presidential elections \citep{van2019influence}. 
%\citep{walsh2018thank,kidder2016college}
Related work has proposed critiques of candidates' use of identity politics \citep{fukuyama2018against,hobsbawm1996identity} 
%and defenses \citep{abrams2019pluribus}
and inspected aspects of how identity interacts with ideology \citep{grossmann2015ideological,kidder2016college}. 
Still more work has studied political advertising in particular, exploring how ads lean into identities or issues \citep{kaplan2006dialogue,latimer1984policy}. 
For example, there is strong evidence that issues influence voters' decision in senate elections \citep{wright1986candidates}. 
Other work has studied the linguistic styles  \citep{jordan20182016,savoy2018analysis,bronstein2013like} or personality traits of candidates \citep{glasgow2000uncertainty}, but not related these to messaging around identities or issues. 
A study of how the spouses of presidential candidates speak found that that they draw on themes of nationality and class \citep{gleibs2018identity}, an identity-driven framing.

Social media has been found to be an important factor in electoral success and campaign strategy \citep{buccoliero2020twitter,fulgoni2016power}, 
although there is ongoing debate about the effectiveness of online interventions on different political behaviors \citep{hersh2015hacking,sides2014gamble}.
It may be that online advertising does not have immediate effects but can strengthen partisan ties \citep{lavigne2021strengthening}.
A study of microtargeted ads on Facebook found that they were largely unpredictive of turnout, although ads that spoke about abortion rights and women's healthcare were potentially effective at mobilizing female voters \citep{haenschen2023conditional}.

Rather than study the effectiveness of candidate messaging on social media, others have described their behavior. One study of Clinton and Trump on Twitter in 2016 found that Trump attacked Clinton and Democrats in 44\% of his tweets and mentioned issues in only 2.3\%, while Clinton made attacks and mentioned issues roughly equally (28\% and 27\% respectively) \citep{buccoliero2020twitter}. This study of the 2016 candidates on Twitter is similar to ours in that it explicitly quantifies the mention of issues.
Furthermore, 
a study comparing TV ads to online ads found that candidates mention issues equally often in each medium; however, they did not explicitly study social media \citep{franz2020issue}.
\citet{stromer2021political} found that Biden and Trump focused on issues more than character in 2020, but did not discuss social group identities. 
Others have found that ads on Facebook are less issue-focused and more partisan than those on TV \citep{fowler2021political}. 
To the best of our knowledge, no one has systematically studied the relative use of identities and issues on social media as we do here.

\subsection{Identities vs. Issues - Expressive and instrumental views of politics}
There is a wealth of research into the question of how to explain party ID, 
both in terms of what drives voters to attach to a given party and in how political candidates market themselves.
Some have argued that 
party identification is formed early in life, such that later attitudes 
and behaviors justify this existing identification (rather than the other way around)
\citep{bartels2002beyond,americanvoter,donald2002partisan}.
The early-in-life model may be compelling, yet it does not provide explanations for how voters may respond to changes in how a party messages about issues or social groups. 
One may choose which groups to incorporate into their political identity over their lifetime \citep{pinsof2023strange}, and some identities, such as one's education level, are acquired over time and not necessarily from one's family. 
A second model (the expressive view) emphasizes social identities \citep{huddy2017political}, in part reinforced by increased social sorting \citep{brown2021measurement,davis2016sorting}.
Generally, 
rather than considering the range of policies 
a certain party promotes, 
a voter can consider the range of social groups
that the party represents, where stronger affiliations with groups lead to stronger affiliations to parties \citep{mason2018one}. 
\citet{kane2021s} find evidence that party choice can even be influenced by a favorable perception of the groups a party is perceived as supporting even if one does not belong to those groups \citep{kane2021s}. 
While this explanation seems vulnerable to the fact that both parties look similar on average, 
\citet{ahler2018parties} found that respondents overestimated the proportion of a party's voters from 
stereotypical groups, leading to the view that social bases of the groups are more different than they are. 
A third model describes a rational voter who 
keeps a running tally of policy accomplishments and rewards the party 
whose successes best reflect their values \citep{fiorina1976voting,achen1992social,lenz2012follow,highton2011long}, 
whether cultural, economic, or both (the instrumental view) \citep{stoetzer2020vote}. An extreme example of this is a single issue voter, 
who chooses their party based on its perceived position on a single issue in which they have  high interest  \citep{congleton1991information}.
Finally, many have proposed models that combine elements of identities and issues \citep{americanvoter,lewis2008american}, 
and the whole framing of choosing a party may best be expressed as a Bayesian process of updating where one begins with  socialized preferences and gradually updates these with more information \citep{achen1992social}.

Regardless of whether identities or issues best describe voters' preferences, or whether it is necessary to model political behavior as being influenced by both, it is also an open question how elites signal to voters through identities and issues. 
While previous work has studied both how different aspects of social identities and issue preferences influence political preferences and participation, there is not a large analogous body of work studying elites, although there are studies on specific parties and identities/issues and political campaigns, with more work focusing on issues and recent work studying the prevalence and effectiveness of negative messaging \citep{fridkin2004negative,lau1999effects}.  We consider this a critical oversight, as  discussions of identities, issues, and negative messaging are prevalent in political science and related areas, understanding the ways in which they present and interact on social media can provide new insight into these ongoing discussions.

\subsection{Negative Advertising} One type of messaging that has been extensively studied is how candidates employ negative emotions \citep{brader2005striking} and attack other candidates \citep{lau1999effects}, both in traditional media \citep{lau1999effects} and 
social media settings \citep{fine2023negativity,auter2016negative,de2012social}. 

While there is ample evidence about the ability of emotional signaling to affect political behavior \citep{brader2005striking}, there is ongoing debate about what role negative messages serve \citep{clinton2004targeted}.
There is evidence  that they are effective at dissuading voters from the candidate 
being attacked \citep{ansolabehere1995going}. 
However, there is also evidence that this strategy is  \textit{in}effective \citep{lau1999effects,clinton2004targeted},
or that it works in some cases \citep{fridkin2004negative}. 
Still, the fact that campaigns believe this strategy works is demonstrated in their tendency to use this method, with 55\% of the ads in the 2016 presidential election employing a negative strategy \citep{galasso2023positive}.
Our primary focus is not on differentiating positive and negative posts, avoiding shortcomings with that categorization \citep{jamieson2000eliminate}. Instead, we are interested in how attack ads are used with identities and issues.

We are not the first to study negative ads on social media. 
For example, 
 the use of negative ads were found to differ across Facebook and Twitter, and across the 2016 and 2020 elections, 
with the 2016 election being remarkably negative \citep{stromer2021political}.
In order to understand how negative messaging relates to how candidates mention identities and issues, we also code each message in terms of whether it disparages the opposition candidate, and when this content mentions identities or issues. 
To the best of our knowledge, no one has studied how attack ads are used with identities and issues.

\subsection{What outcome are candidates targeting on TikTok late in the election cycle?}
Political communication can be earned (i.e., organic) or paid (i.e., advertisement) and it could be categorized as aiming for one of three key goals (stylistically): persuasion (aimed at people who are likely to vote, but not sure who they are going to vote for), get-out-the-vote (aimed at people likely to support the candidate, but not sure if they are going to vote), and fundraising (aimed at strong supporters who are already likely to vote for the candidate).

Research done in conjunction with Facebook for the 2020 election showed the vast majority of advertising content by the major campaigns was aimed towards fundraising \citep{allcott2025effects}. Others have found that different calls to action occur on different platforms \citep{larsson2024calls,wurst2023mobilization} and at different times of the election cycle, with later stages focusing on getting out the vote and fundraising \citep{stromer2021political}. And, organic content on social media is also aimed at co-partisans who represent the vast majority of followers of partisan accounts \citep{yu2024partisanship}. Thus, whether fundraising or get-out-the-vote, it is likely that the vast majority of the content late in the cycle from candidate accounts is aimed towards co-partisans.

\section{Study Design and Data}
Our study aims to understand if and how candidates speak to political identities and issues on the relatively young social media platform TikTok. To understand the implications of their posting strategies, we inspect how they align with the opinions of the American public. That is, do candidates speak to the identities/issues that are important to members of their party, or do they speak to those which best differentiate them from the other candidate?
Answering our questions requires both utilizing social media and the collection of novel survey data. 
To ensure that we are understanding how candidates are informed by public opinion and not the other way around, we design and implement our survey far before the 2024 presidential election, in the winter of 2022. 

\subsection{Social media design and data}
We would like to see the campaign posts that an average TikTok user might see on their For You Page. 
Towards this end, we create approximately 200 sock puppet accounts and record all of the posts they see via simulated browser-based interaction with the For You Page \citep{sandvig2014auditing}. 
These accounts log onto TikTok twice per day for a thirty-minute period - between 7:00-9:00 am and 7:00-9:00 pm. This provides a view of the types of content the For You Page algorithm \citep{klug2021trick} is likely to show to a typical user.

These accounts were grouped into four quadrants, which we outline briefly here. 
\textbf{Quadrant 1 - Lifestyle-heavy:}
In this quadrant, the majority of the channels each account follows are lifestyle influencers and not news accounts.
\textbf{Quadrant 2 - Balanced:}
In this quadrant, each account follows the same number of lifestyle and news accounts. 
\textbf{Quadrant 3 - Disengaged:}
In this quadrant, accounts do not follow any accounts. 
\textbf{Quadrant 4 - News-heavy:}
In this quadrant, each account follows a majority of news accounts.
All accounts were logged on through the AWS Ohio region. We choose this region as it is a region in the Midwest of the United States where both candidates are likely to post on TikTok. The main dataset is all of the posts that are from campaign or campaign-adjacent accounts (kamalaharris, kamalahq, timwalz, realdonaldtrump, teamtrump, and jd). As shown in \tabref{tiktok} , during July 28th, 2024 - November 22nd, 2024, we saw 518 posts posted by the Harris (304 posts) and Trump (214 posts) campaigns\footnote{As we see that the 6 posts by the realdonaldtrump channel which were not seen by the sock puppets were very popular, we also repeat the primary TikTok analysis on these posts in \secref{sec:robustness}. We see that were we to include these 6 posts our results would not qualitatively change.}.

\begin{table}[]
\begin{adjustbox}{width=\columnwidth,center}{
\begin{tabular}{lrrrrr}
\toprule
                      & Team/Account         & \#Unique & Median views   & Median likes   & Median comments \\
                      & \multicolumn{1}{l}{} & posts   & (in thousands) & (in thousands) & (in thousands)  \\ \midrule
Campaign team         & Harris               & 304/791  & 6400/1000      & 966/103        & 8/1             \\
                      & Trump                & 214/309  & 9900/1200      & 1000/106       & 21/3            \\ \midrule
Harris team breakdown & kamalaharris         & 82/137   & 3250/1600      & 430/132        & 9/4             \\
                      & kamalahq             & 204/570  & 1600/960       & 194/98         & 2/1             \\
                      & timwalz              & 18/84    & 1450/538       & 256/76         & 4/2             \\ \midrule
Trump team breakdown  & realdonaldtrump      & 49/6     & 7500/10950     & 773/602        & 19/22           \\
                      & teamtrump            & 149/281  & 4350/1200      & 401/93         & 10/3            \\
                      & jd                   & 16/22    & 1500/632       & 227/84         & 6/2            
 \\ \bottomrule 
     
\end{tabular}
}
\end{adjustbox}
\caption{Data are shown in the format ``posts in/not in analysis''. TikTok data from both campaigns July 28th, 2024 to November 22nd, 2024.
We analyze the posts collected through sock puppets.  This provides a view into the types of posts a non-politically engaged typical voter may see on TikTok. Here we compare the posts in the analysis to those not in the analysis. Those not in the analysis are simply posts on the page of each associated account which we inspected manually by visiting each page. 
We can see that those posts in the analysis are more popular, and have more views, likes, and comments, compared to the posts by the same accounts which are not viewed by the sock puppets. }
\label{tiktok}
\end{table}

 The lifestyle influencers are randomly picked from a pool of accounts listed in Appendix Table \ref{tab:influencers}, while news channels are randomly picked from the accounts listed in Appendix Table \ref{tab:news}:

\begin{itemize}
    \item Quadrant 1 (lifestyle-heavy): accounts follow 10 lifestyle influencers and 2 news channels.
    \item Quadrant 2 (balanced): accounts follow 6 lifestyle influencers and 6 news channels.
    \item Quadrant 3 (unengaged): accounts follow no one.
    \item Quadrant 4 (news-heavy): accounts follow 2 lifestyle influencers and 10 news channels.
\end{itemize}

We created and maintained 49 accounts in Quadrant 1, 48 accounts in Quadrant 2, 55 accounts in Quadrant 3, and 49 accounts in Quadrant 4. We see rudimentary evidence that TikTok recommends campaign posts to all quadrants because of their popularity, yet quadrants that follow more news channels see more campaign posts and the sock puppets were shown different posts, presumably tied to their different seeding. Specifically, accounts in Quadrant 1 (lifestyle-heavy) see a total of 130 unique campaign posts, while this number for Quadrant 2 (balanced) is 156, for Quadrant 3 (disengaged) is 133, and for Quadrant 4 (news-heavy) is 208. On average, accounts in Quadrant 1 see 9.55 campaign posts (SE=0.723) during our data collection, while this number for Quadrant 2 is 13.06 (SE=0.811), for Quadrant 3 is 15.13 (SE=0.751), and for Quadrant 4 is 17.98 (SE=1.002).

\subsection{Survey design and construction of relevant models and variables}
We deployed two surveys using the firm Prodege and the software platform Qualtrics. 
The first was launched in Winter 2022 and the respondents who successfully completed 
the survey were recontacted one month later. 
The first survey contained a battery of questions around social group identity, 
whereas the second focused on issue positions. These two surveys are included in 
the appendix (see Appendix Section \ref{sec:survey_doc}). We received 1,997 responses to the first survey and 1,027 of these 
successfully completed the second round. 
The population was contacted following a quota to match the adult population\footnote{Quotas were: gender (max 1,200 for either gender), age (max 600 for three age groups, but 800 for 45-64), race (max 1,440 for white, max 400 for Black, and max 600 for Hispanic. Here, we use the term Hispanic as an aggregate label for respondents who identified as Hispanic, Mexican, Cuban, Puerto Rican, or with another Hispanic group which they wrote in a free response text box.), education (max 800 for high school or less, max 600 for some college, max 800 for college or more), and party max 750 for either party or other (on a 7-point scale).} 
of the US on gender, age, education, race, and partisanship\footnote{This research was deemed exempt by the authors' IRB.}.

\begin{figure}
    \centering
    \includegraphics[width=\linewidth]{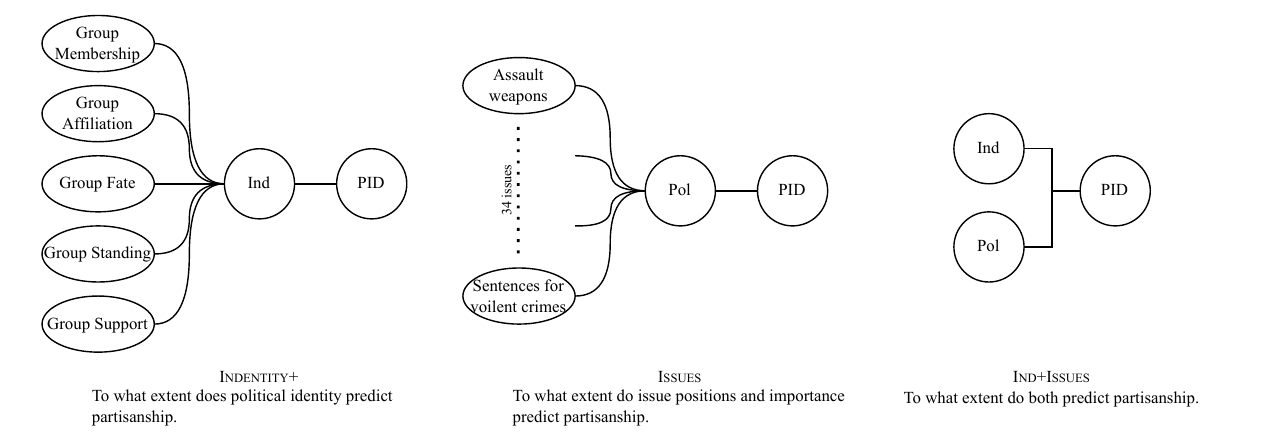}
    \caption{We explore three models to predict partisanship. Each model involves some combination of items on identity and items on political issues.}
  \label{models}
\end{figure}

 To understand how identities and issues predict party ID we form three models (shown in  \figref{models})
 using the survey responses. 
The first, \modelone, 
models partisanship as a function of one's identities. 
As identity is a difficult-to-measure construct with competing definitions, we take a holistic approach and collect measures of five key variables: Group Membership, Group Association, Group Fate, Group Standing, and Group Support. Additional details around the construction of these variables are included in \secref{sec:extra} (Additional methods background and details).

\modeltwo, models partisanship as a function of  issue position and importance \citep{fournier2003issue}, while \modelthree, models partisanship as a function of both identities and issues, as  identities and issue positions likely both influence 
one's partisanship.
We model partisanship with a seven-point scale and our final outcome variable is grouped into three buckets: Democrat (Strong Democrat, Weak Democrat, Leans Democrat), Republican (Strong Republican, Weak Republican, Leans Republican), or Independent (Neither).
To mitigate potential bias, this is the final question that respondents see in the section on demographics in the first survey. 
We wait one month before asking respondents to take the survey on policy issues so that their responses to the identity questions do not bias their responses to the issue questions. 

\section{Analysis}
Here we address the three research questions from above, investigating the extent to which presidential candidates spoke about identities and issues, which of these they discussed, and their affect.
Then, to understand how the identities and issues discussed by candidates aligned with the attitudes of potential voters within their political party, we consider a set of sub-questions. 
We begin by describing the identities and issues of Democrats and Republicans, so that we can compare their positions to the messages of politicians. Next, we investigate which identities and issues are important to each group. 
We then use the survey responses to build predictive models, in order to understand if identities or issues are more predictive of party ID, and which particular identities and issues are predictive. 
These sub-questions, along with the analysis in response to RQ1 and RQ2, allow us to understand if candidates speak to the identities and issues that are descriptive and/or discriminative, and broadly if their strategy aligns with the attitudes of potential voters. 

We include robustness checks and other additional analyses in the Appendix Section \ref{sec:add_analysis}.

\subsection{RQ1: Did the presidential candidates in the United States 2024 election speak to identities or issues? Which did they mention more? And, what was the interaction between identities and issues, and negative affect?}

\begin{figure}[hb!]
    \centering
    \includegraphics[width=\linewidth]{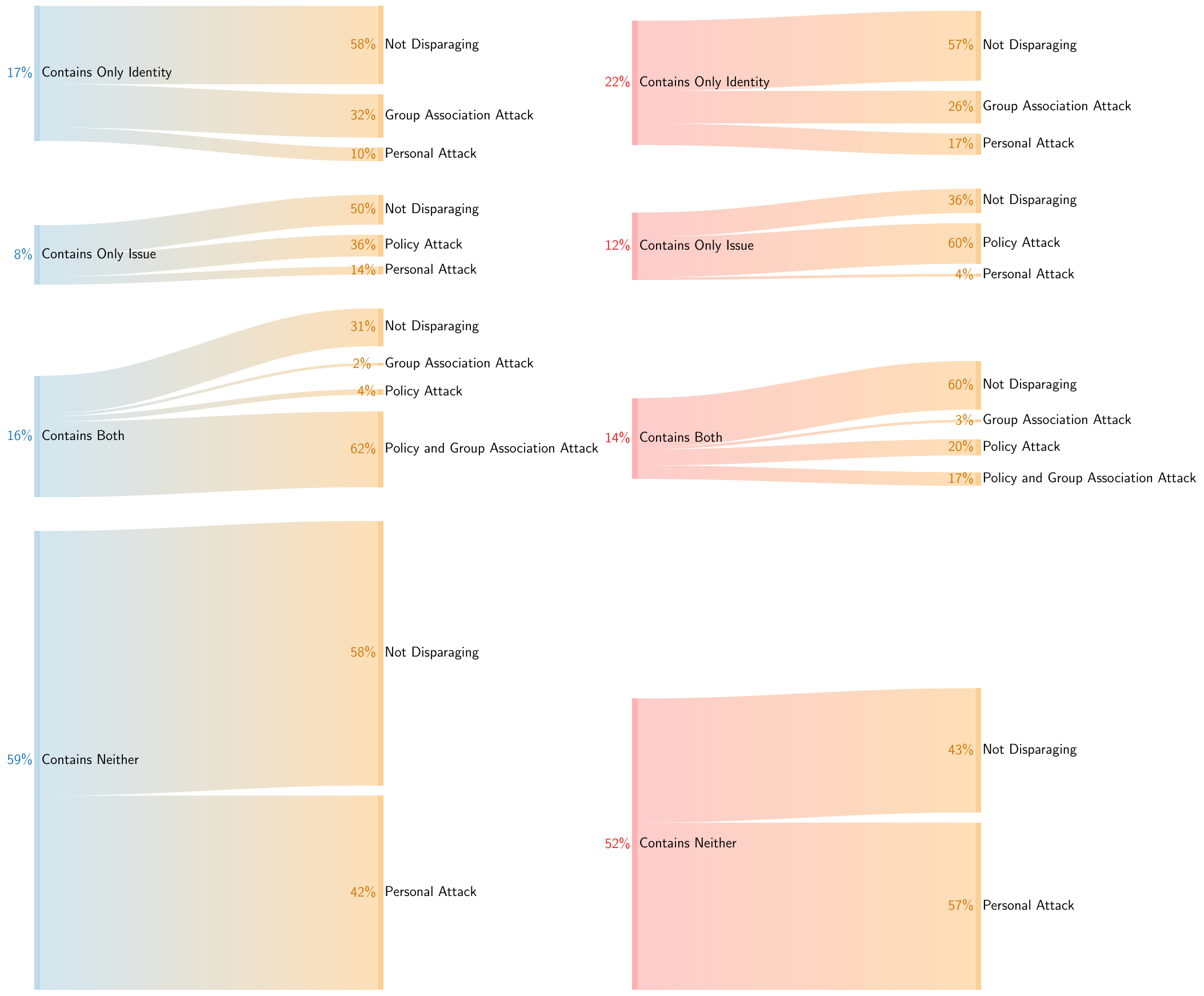}
    \caption{ Harris (left) and Trump (right) TikTok posts in terms of containing identity, issue, and disparaging content. We see that 50\% of Harris posts and 54\% of Trump posts contain disparaging content. Among all Harris posts, 12\% are policy attacks, 15\% are group association attacks, and 26\% are personal attacks. Among all Trump posts, 12\% are policy attacks, 8\% are group association attacks, and 33\% are personal attacks. The width of nodes represents the number of posts. }
    \label{fig:basics}
\end{figure}

We begin by inspecting how often each candidate discusses identities and issues. 
In \figref{fig:basics}, we see that both candidates mention identities more than issues.
Harris mentions identities in 33\% of her posts and issues in 24\%, whereas Trump mentions identities 
in 36\% and issues in 26\%. Trump is slightly more likely to mention identities than Harris. 
However, what is most striking about \figref{fig:basics}
is that both candidates are most likely to discuss neither identities nor issues.

We code all campaign posts seen by the accounts to match the identity and policy issue categories of our survey design. We also code campaign posts on whether they contain disparaging content toward the opposing party or candidate. Here, we consider four categories for disparaging content: Policy Attack, Group Association Attack, Personal Attack, and Not Disparaging. Policy Attack describes posts that attack Trump or Harris for a policy action, proposal, or just association with some kind of policy. Group Association Attack describes posts that attack Trump or Harris for either failing to support a certain group (e.g. an attack claiming that Trump is prejudiced against Latinos) or supporting a group (e.g. an attack claiming that Harris only cares about transgender rights). Personal Attack describes posts that attack Trump or Harris solely based on their unsuitability or incompetence. If a post (1) does not attack the other candidate personally, and (2) either doesn't state the policy or identity preference of the other candidate or states the facts of their campaign portfolio matter-of-factly, the post is labeled as ``Not Disparaging''. All posts have a label in this system.

Harris' team disparages Trump in 50\% of their posts, while Trump's team disparages Harris in 54\% of their posts. 
The fact that we see so many disparaging posts speaks to a general trend of negative campaign ads \citep{geer2012news} and communication trends of U.S. Representatives and Senators \citep{westwood2024pulse}. Among Team Harris posts, 26\% contain Personal Attacks, 15\% contain Group Association Attacks, and 12\% contain Policy Attacks. Among Team Trump posts, 33\% contain Personal Attacks, 8\% contain Group Association Attacks, and 12\% contain Policy Attacks. For examples of these attacks, see Table \ref{tab:exampleQuotes}. 

Overall, both candidates issue more Personal Attacks than any other category. Harris is less likely to issue Personal Attacks in posts that mention neither identities nor issues. In these posts, she introduces herself as a candidate with positive messages. 
%htbp
\begin{table}[!htbp]
\centering
\adjustbox{max width=.8\linewidth}{%
\begin{tabular}{lll p{2cm} p{2cm}}
\toprule
Team   & Type of Attack            & Quote                                                 & Account         & Date       \\ \midrule

Trump  & Personal           & \begin{tabular}[c]{@{}l@{}}The video has a clip from CNN where the \\host asked a boy ``What's the first word that \\pops up your head when you hear the name \\Kamala Harris?'' To which the kid answered, \\``Liar.'' And the video description says ``So \\true, kid!'' \end{tabular}                                                                                                                                                                                     & teamtrump       & 2024-9-26 \\         &                   & \multicolumn{3}{l}{https://www.tiktok.com/@teamtrump/video/7418975389944401198} \\ \\ \addlinespace
       & Group Association  & \begin{tabular}[c]{@{}l@{}}The video started with a clip from Harris' \\ earlier videos where she said, ``18 through \\ 24, they are stupid". Then, Trump said, \\ ``She's not the brightest bulb in the ceiling. \\ No, she's not smart. We're with the young \\ people. They are smart. We love the young \\ people, and we're gonna take them every \\ time."\end{tabular} & realdonaldtrump & 2024-7-27  \\         &                   & \multicolumn{3}{l}{https://www.tiktok.com/@realdonaldtrump/video/7396435731117280543} \\ \\ \addlinespace
       & Policy             & \begin{tabular}[c]{@{}l@{}}``They've had three and a half years to fix \\ the border. They've had three and a half \\ years to create jobs. And all the things we \\ talked about. The worst president, the \\ worst vice president in the history of our \\ country."\end{tabular}                                                                                          & realdonaldtrump & 2024-9-14 \\          &                   & \multicolumn{3}{l}{https://www.tiktok.com/@realdonaldtrump/video/7414599502465600798}\\ \midrule
Harris & Personal           & \begin{tabular}[c]{@{}l@{}}``Tim Walz on Trump and Vance: \\ `These guys are weird as hell.'"\end{tabular}                                                                                                                                                                                                                                                                   & kamalahq        & 2024-8-7   \\        &                   & \multicolumn{3}{l}{https://www.tiktok.com/@kamalahq/video/7400522330935414047}\\ \\ \addlinespace
       & Group Association  & \begin{tabular}[c]{@{}l@{}}The video has a clip from one of Trump's \\ rallies, where a speaker said on the stage, \\ ``These Latinos, they love making babies."\end{tabular}                                                                                                                                                                                                & kamalahq        & 2024-10-28 \\  &                   & \multicolumn{3}{l}{https://www.tiktok.com/@kamalahq/video/7430918268728216875} \\ \\ \addlinespace
       & Policy             & \begin{tabular}[c]{@{}l@{}}``Donald Trump hand-selected three \\ members of the Supreme Court with \\ the intention that they would undo the \\ protections of Roe v. Wade, and they \\ did exactly as he intended."\end{tabular}                                                                & kamalaharris    & 2024-9-10 
 \\  &                   & \multicolumn{3}{l}{https://www.tiktok.com/@kamalaharris/video/7413200952662035742} \\ \bottomrule
\end{tabular}}
\caption{Examples of different types of attacks found in TikTok campaign posts.}
\label{tab:exampleQuotes}
\end{table}

% \begin{figure}
%     \centering
%     \includegraphics[width=0.95\linewidth]{smfigs/basics.pdf}
%     \caption{Caption}
%     \label{fig:basics}
% \end{figure}

% \begin{figure}
%     \centering
%     \includegraphics[width=0.95\linewidth]{smfigs/dis_content.pdf}
%     \caption{Caption}
%     \label{fig:diss}
% \end{figure}

\newpage

\newpage
\subsection{RQ2: Which identities and issues did they discuss?}
\begin{figure}[htb]
    \centering
    \includegraphics[width=.9\linewidth]{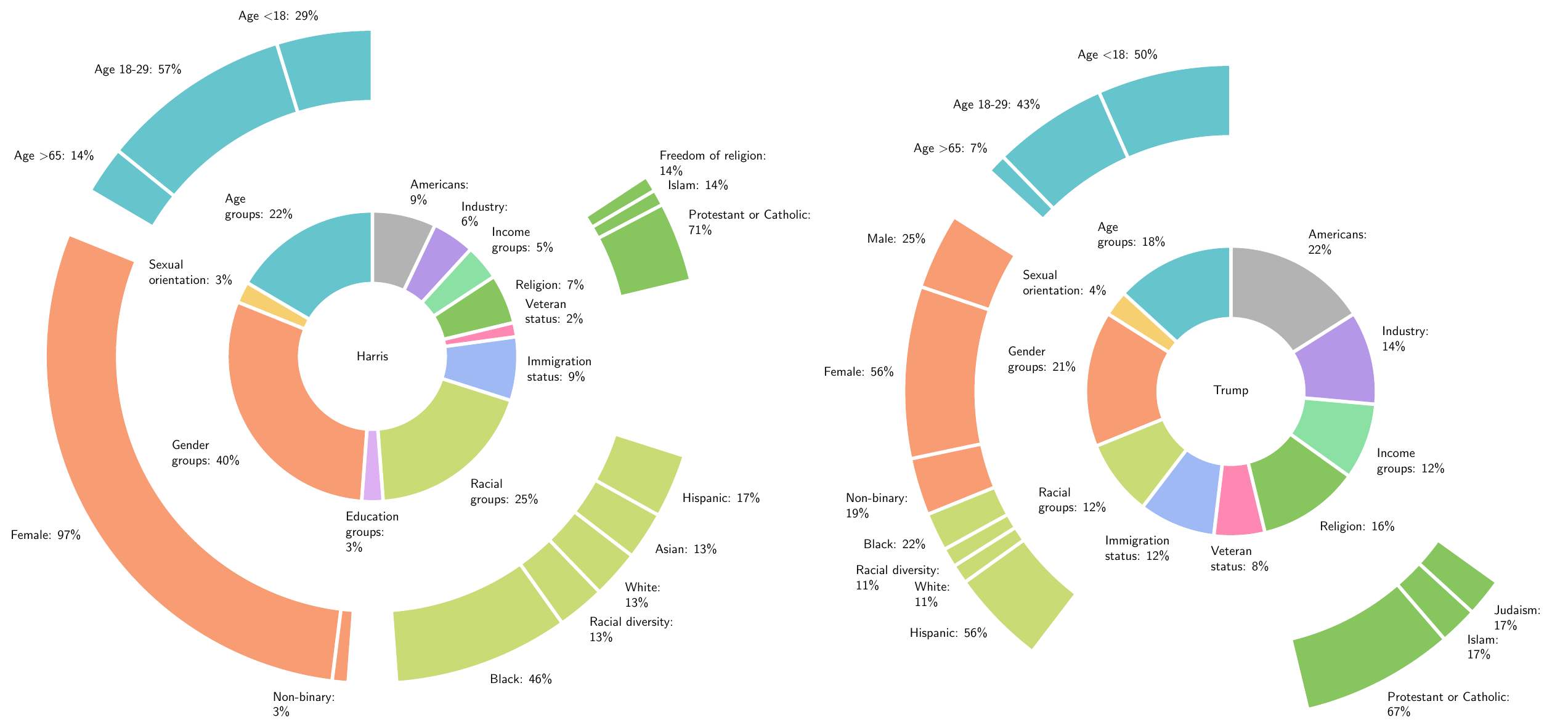}
    \caption{ The social identities mentioned by Harris and Trump within TikTok posts. Here, we tagged explicit mentions and did not tag a group if it was referred to implicitly (e.g. by the camera panning over members of that group). }
    \label{fig:smids}
\end{figure}

As both candidates speak about identities to some extent, 
Harris in 33\% of posts and Trump in 36\% of posts, 
we next inspect the particular identities they discuss in \figref{fig:smids}.
We see that Harris most frequently discusses 
gender groups, followed by racial groups, age groups, and immigration groups. 
In total Harris mentions 11 of these groups, 
whereas Trump mentions 10. 
In \figref{fig:smids}, we also see that Trump 
mentions ``Americans'' the most, followed by gender and religious groups. 
In this dataset, Harris mentions education groups, 
whereas Trump does not.

We see that both candidates mention females more than males. 
However, they do this in different ways when they mention females without explicitly bringing up issues. 
Harris is most likely to mention females through a frame of rights. 
For example, she often says she does not want women today to have fewer rights than women of previous generations. 
Trump tends to portray females through a male lens, as needing protection. 
However, they both bring up females in connection to issues; 
Harris in connection to reproductive rights, and Trump in connection to punitive actions against women pursuing abortion, transgender athletes, and protection from immigrants.

% \begin{figure}
%     \centering
%     \includegraphics[width=0.95\linewidth]{smfigs/harris_ids.pdf}
%     \caption{Harris}
%     \label{fig:smids_harris}
% \end{figure}

% \begin{figure}
%     \centering
%     \includegraphics[width=0.95\linewidth]{smfigs/trump_ids.pdf}
%     \caption{Trump}
%     \label{fig:smids_trump}
% \end{figure}

% \begin{figure}
%     \centering
%     \includegraphics[width=0.95\linewidth]{smfigs/harris_issues.pdf}
%     \caption{Harris}
%     \label{fig:issues_harris}
% \end{figure}

\begin{figure}[htb]
    \centering
    \includegraphics[width=1\linewidth]{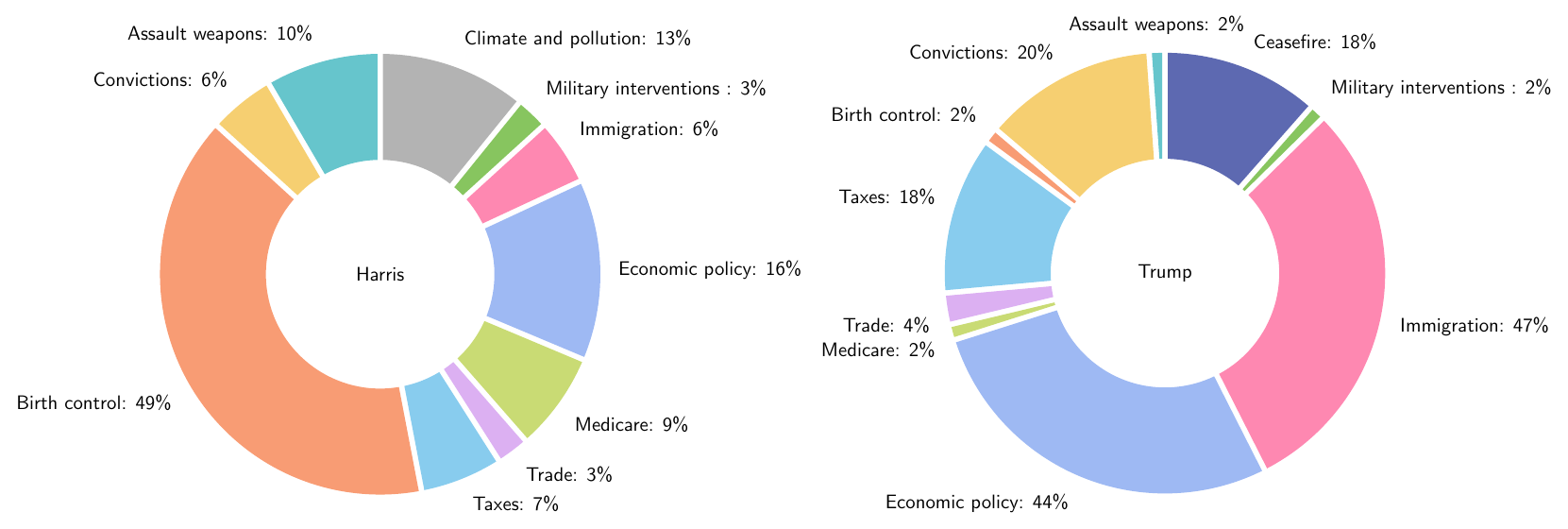}
    \caption{ The policy issues mentioned by Harris and Trump within TikTok posts. Unlike social identities, here we do not break the issue categories into more granular issues as the candidates do not often discuss these issues in sufficient detail to do so. }
    \label{fig:issues}
\end{figure}

We also investigate the particular issues that the candidates discuss. 
We see that 
Harris discusses issues in 24\% of posts and Trump in 26\% of posts, 
we next inspect the particular issues they discuss in \figref{fig:issues}.
In \figref{fig:issues} we see that Harris most frequently discusses 
birth control, economic policy, and climate and pollution. 
Both Trump and Harris mention 10 issue categories. Yet, Harris discusses climate and pollution and Trump does not, and Trump discusses ceasefire and Harris does not.
In \figref{fig:issues} we also see that Trump 
mentions immigration policy the most, followed by economic policy, and convictions.

\subsection{RQ3: How do the identities and issues discussed by candidates align with
the attitudes of potential voters within their political party?}

To describe alignment between the candidates and potential voters, we conducted a two-wave survey in 2022. 
First, respondents were asked a series of questions pertaining to their social identities. 
At the end of this survey was a seven-point question about party identification (party ID). 
In the second survey, we asked a series of questions about political issues. 
Using these surveys, we break this research question into several composite questions, before synthesizing them in response to how campaign videos align with the attitudes of potential voters: 

\begin{itemize}
    \item (RQ3 a) What are the identities and issue positions of Democrats and Republicans?
        \item (RQ3 b) Which identities and issues are most important to each political party?
    \item (RQ3 c) Is party ID better predicted by identities or issues?
    \item (RQ3 d) Which identities and issues are most predictive of party ID?
\end{itemize}

The first question allows us to understand the landscape of potential voter attitudes. 
With the second question, we inspect whether candidates are appropriately speaking to the identities and issues which describe the potential voters within their party. The third question is critical to understanding alignment between candidates behavior, and the opinions of potential voters. 
In the TikTok data, we found that candidates 
discussed identities more than issues, suggesting that candidates may believe that voters follow an expressive model of politics, and care about the extent to which a given party represents the identities which are important to them.
Now, we can see if identities better predict which party a respondent reports belonging to. If they do, this suggests that candidates are appropriately messaging to identities rather than issues. 
Alternatively, with the fourth question, we inspect if candidates are  speaking to the identities and issues which \textbf{differentially} describe  the potential voters within their party.

\subsubsection{RQ3 a: What are the identities and issue positions of Democrats and Republicans? }

\begin{figure}[H]
    \centering
    \includegraphics[width=.69\linewidth]{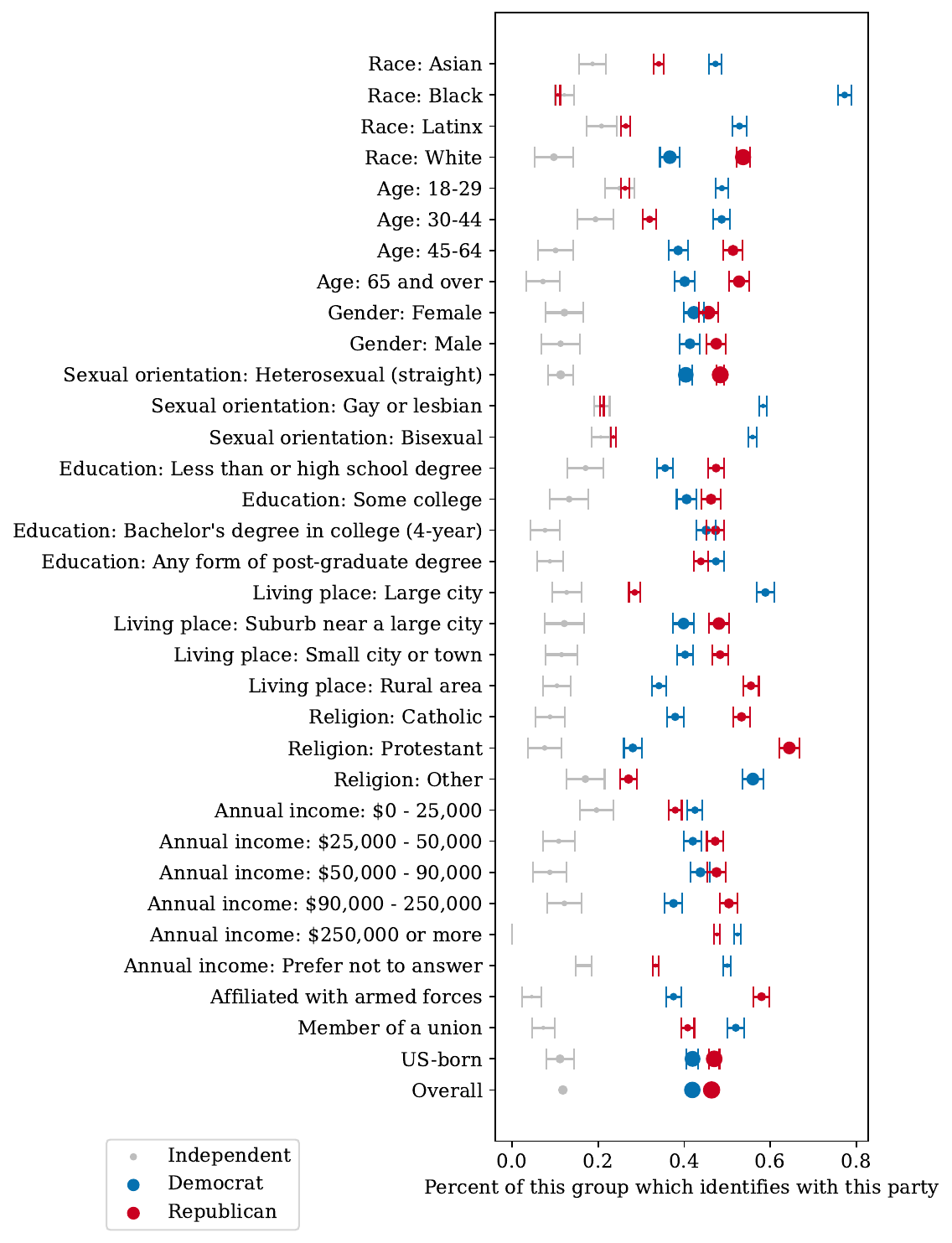}
    \caption{ We show the percentage of respondents who self-described as Democrat, Republican, or Independent, which can be described with each demographic or group membership variable. }
    \label{fig:description_ids}
\end{figure}

We first inspect the dataset in terms of identities and issues in order to answer the questions above. 
In \figref{fig:description_ids}, we see that Protestant (64.42\% of this group), affiliated with armed forces (57.95\%), and rural (55.49\%) respondents are most likely to be Republican. In contrast, Black (77.27\%), urban (58.86\%), and LGBTQ (58.33\%) respondents are most likely to be Democrats. 

\begin{figure}[H]
    \centering
    \includegraphics[width=.9\linewidth]{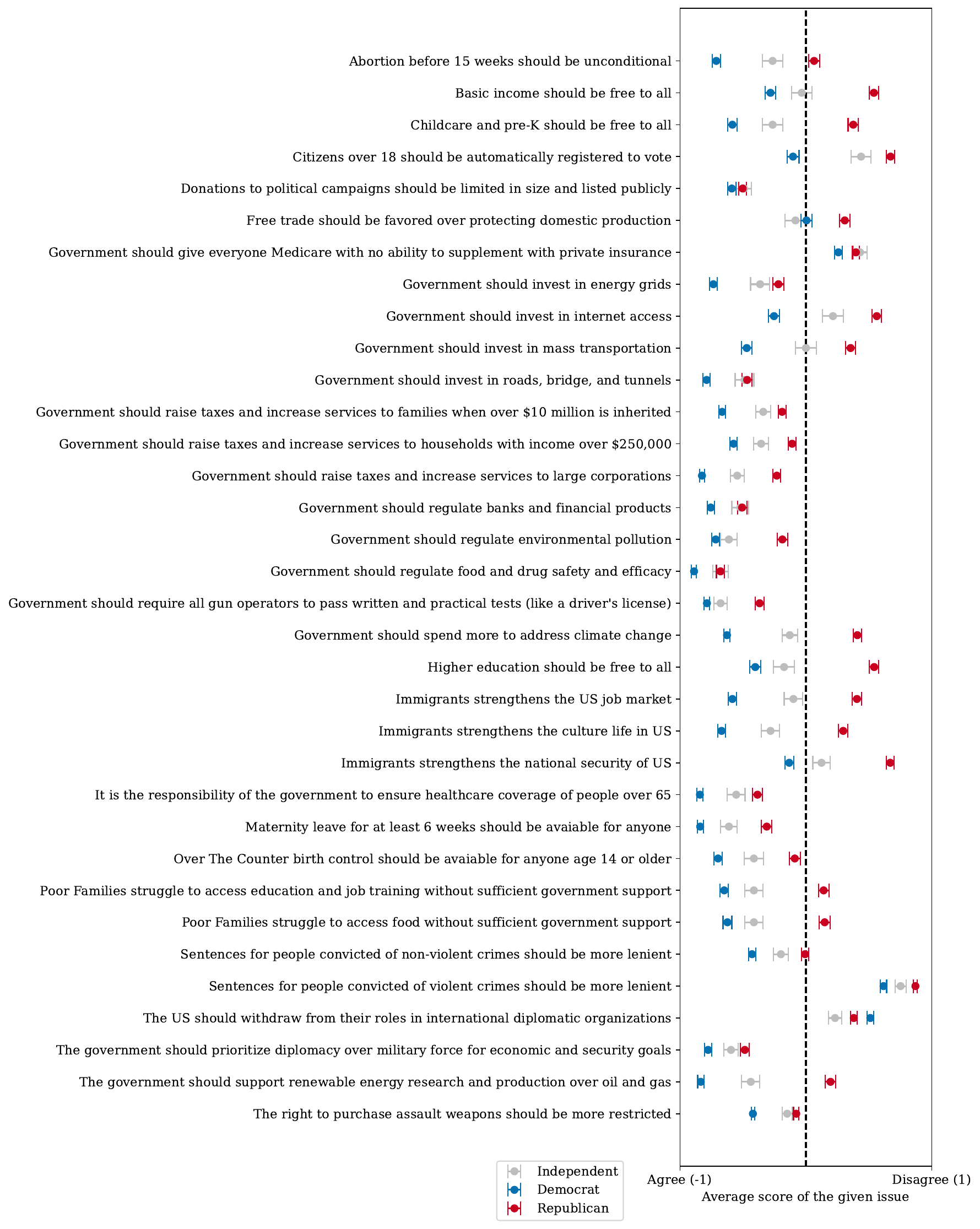}
    \caption{ We show the average issue position from Agree (coded as -1) to Disagree (coded as 1) for respondents who self-described as Democrat, Republican, or Independent.}
    \label{fig:issues_categories}
\end{figure}

In \figref{fig:issues_categories}, where we code ``Agree'' as -1 and ``Disagree'' as 1, we see that Democrats have strong positions on 
%government regulation of food and drug safety and efficacy (mean=-0.8884),
government-provided healthcare for people over 65 (mean=-0.8419, high agreement), and access to at least six weeks of maternity leave (mean=-0.8395, high agreement), among other issues. Republicans have strong positions on reducing sentences for violent crimes (mean=0.8697, high disagreement), %government regulation of food and drug safety and efficacy (mean=-0.6786), 
and automatically registering citizens to vote (mean=0.6723, high disagreement). 

The positions of those who identify with the parties differ the most when it comes to climate change (Democrat mean=-0.6267, Republican mean=0.4086, diff=1.0354), research and production on renewable energy over oil and gas (Democrat mean=-0.8349, Republican mean=0.1954, diff=1.0303), and the positive impact of immigration the job market (Democrat mean=-0.5837, Republican mean=0.4054, diff=0.9892). 

In contrast, most respondents agree about restricting donations to political campaigns (both parties agree, diff=0.0862), the US withdrawing its role in international diplomatic organizations (both parties disagree, diff=0.1314), and the government giving Medicare with no ability to supplement with private insurance (both parties disagree, diff=0.1389).

%The next question is how the survey results capturing everyday peoples' concerns and priorities align 
%with candidate positioning? 

\subsubsection{RQ3 b: Which identities and issues are most important in each political party?}
% \vspace{-8mm}
To understand which identities are most important to potential voters of each party, we ask respondents a question designed 
for them to rank the groups they belong to in a way which mirrors a view of politics where voters select candidates
who distribute resources to specific groups (described in more detail in the section Additional methods background and details). 
To understand which issues are most important, we ask respondents to assign tokens to issues, a task we design such that not all issues can receive tokens, producing a ranking of issues which respondents regard as important. 

\begin{figure}
    \centering
    \includegraphics[width=0.49\linewidth]{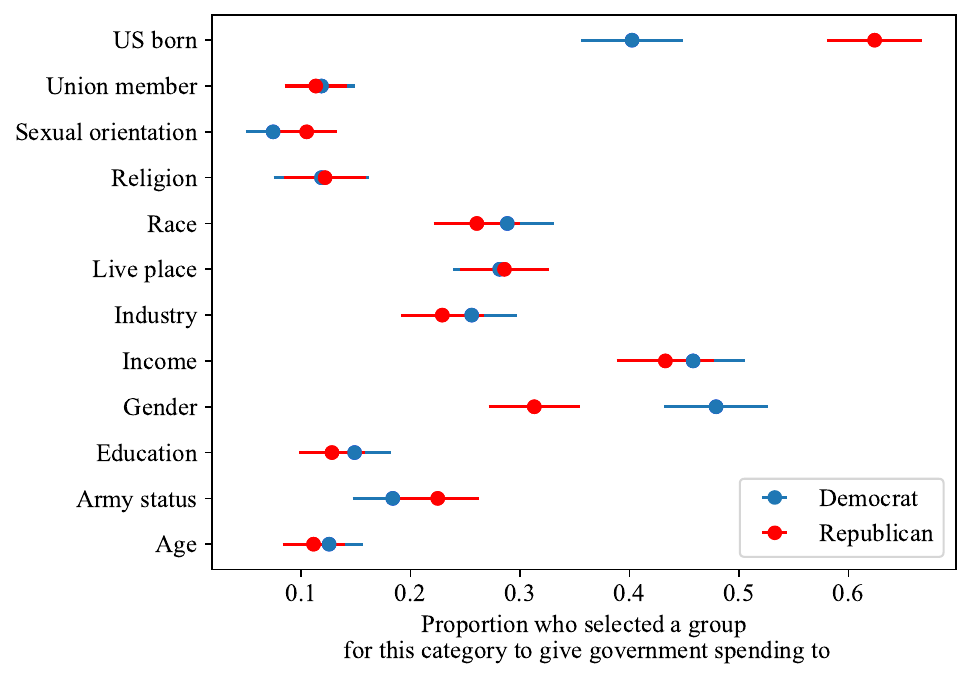}
    \caption{Respondents were asked to select between 1 and 3 groups that they would direct government spending towards. Critically, respondents were only shown particular groups that they said they belonged to. This allows us to use this as a measure of identity (Group Support) as it indicates a measure of relative importance within the groups they belong to. }
    \label{fig:gov_giv}
\end{figure}

In \figref{fig:gov_giv}, we see that there are large differences in the groups that each party would send government spending to. 
For example, we see that a large number of Republicans (60\%) opted to send government spending 
to youth with the same nationality as them, compared to 40\% of Democrats. This is almost entirely spending directed at people born in the US. That is 99\% of Republicans and 96\% of Democrats who would send spending to youth born in the same country as them were born in the US.
In contrast, close to 50\% of Democrats would give to youth of the same gender as them, 
while only 30\% of Republicans would. This giving is largely driven by women, where 87\% of Democrats and 80\% of Republicans who would give to youth of their own gender are women.

\begin{figure}
    \centering
    \includegraphics[width=0.59\linewidth]{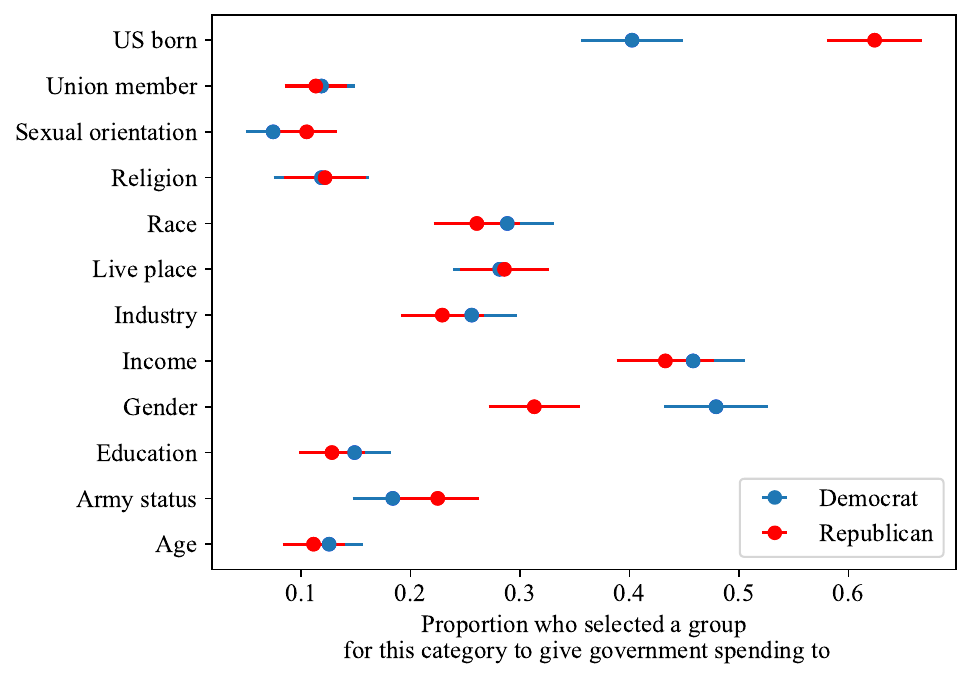}
    \caption{Respondents were asked to assign tokens to the issue categories shown here. They were restricted by the number of issues they could assign tokens to. Here, we lose some specificity by only showing issue categories to respondents; however, we choose this aggregation to reduce the total options participants see. }
    \label{fig:qv}
\end{figure}

In \figref{fig:qv}, we see that Democrats and Republicans largely place different values 
on different issue categories (when asked which policy categories they would want to lobby over). However, Healthcare is a clear exception and is the issue with the greatest importance for potential voters of both parties. The issues with the greatest separation between the parties are 
Environmental Protection and Extraction (which is highly important to Democrats), Taxes (which is highly important to Republicans), and Abortion and Maternity Care (which is highly important to Democrats). 
Next, we inspect which of the combinations of identities and issues are most predictive of party ID.

\subsubsection{RQ3 c: Is party ID better predicted by identities or issues? }
To address this question, we employ a series of models as shown in \figref{models}. Each model utilizes a set of features. The first 
utilizes survey responses around identity, the second around issues, and the third around both. 

Each model is implemented using Logistic Regression. 
 Table \ref{tab:results} shows the predictive performance of each model. We see significant F1 score (the harmonic mean between precision and recall) gains with \modeltwo{} relative to \modelone, 
 and these gains occur both in the precision and recall for both parties. 
In \modelthree, we see that identity information can slightly assist with performance when issue positions are already known. 
%And, we see with \modelfour, that issues are not wholly predictable with a rich identity model.

\begin{table}[h!]
    \centering
    \resizebox{\linewidth}{!}{
\begin{tabular}{llrrrlrrrlrrr}
\toprule
           &  & \multicolumn{3}{c}{Democrats} &  & \multicolumn{3}{c}{Republicans} &  & \multicolumn{3}{c}{Overall}   \\
           &  & Precision & Recall & F1 Score &  & Precision  & Recall  & F1 Score &  & Precision & Recall & F1 Score \\ \cline{3-5} \cline{7-9} \cline{11-13} 
\modelone   &  & 0.573     & 0.793  & 0.665    &  & 0.700      & 0.634   & 0.661    &  & 0.631     & 0.712  & 0.663    \\
\modeltwo   &  & 0.741     & 0.860  & 0.797    &  & 0.779      & 0.864   & 0.819    &  & 0.760     & 0.862  & 0.808    \\
\modelthree &  & 0.743     & 0.868  & 0.800    &  & 0.784      & 0.864   & 0.822    &  & 0.763     & 0.866  & 0.812   \\
\bottomrule
    \end{tabular}
    }
    \caption{ We see significant performance gains when we include issue positions and importance in the predictive models. }
    \label{tab:results}
\end{table}

Thus we see that issues are much more predictive than identities. This is mirrored when we look at the feature importance in the joint model in \figref{fig:features}, where we see that the weight on the issue features is typically greater than that on the identity features.

\subsubsection{RQ3 d: Which identities and issues are most predictive of party ID?}

\begin{figure}[ht!]
    \centering
    \includegraphics[width=.8\columnwidth]{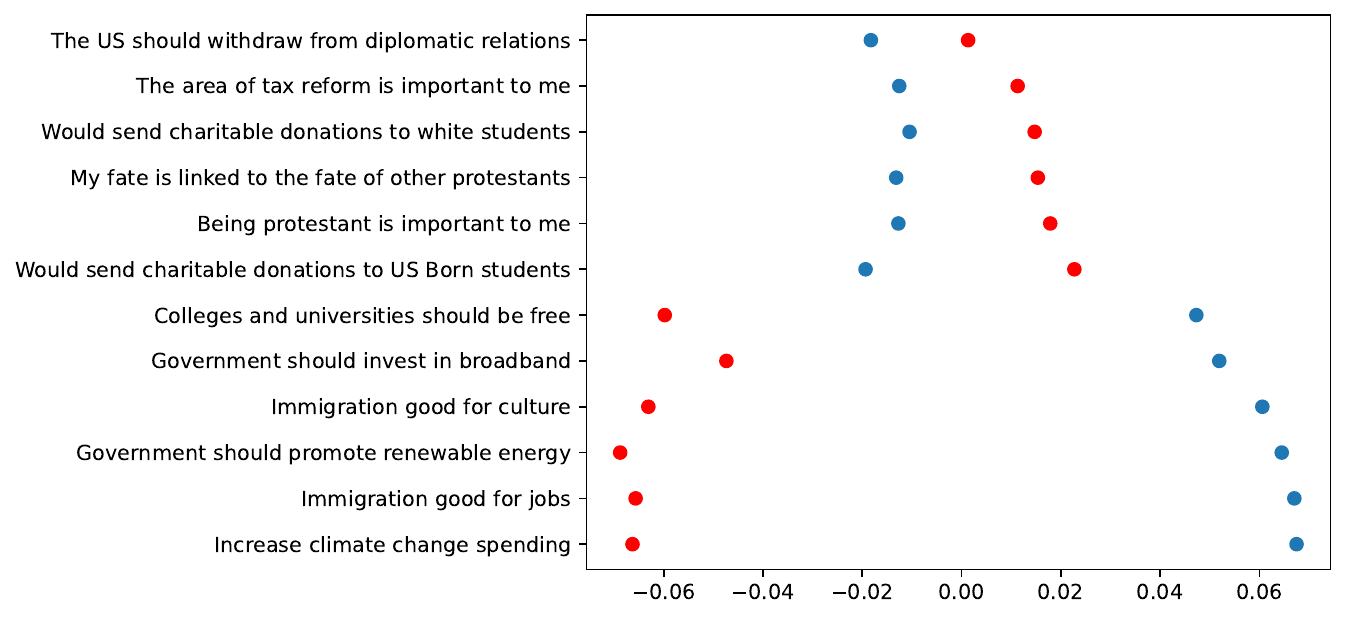}
    \caption{\textbf{The raw coefficient value.}
    Inspecting \modelthree, we show the union of the five most positive and five most negative features for predicting each party.}
    \label{fig:features}
\end{figure}

In Figure \ref{fig:features}, we show the union of the five most positive and five most negative features when predicting each party with \modelthree{}.
The identity categories which are most predictive of party ID are: religion (measured as Group Association and Group Fate with protestants), 
race (measured as Group Support for white people), and nationality (measured as Group Support for US-born students). 
The issues which are most predictive are: climate change / renewable energy spending, immigration, and government spending. 

%We see that a few issues, climate change; immigration; and spending, are highly predictive. Additionally, we see that a few identity features are important as well. 
All of the identity variables align well with research on white identity politics \citep{jardina2019white}, and are predictive of Republican identity. Although the inclusion of these features does not drastically affect the model’s predictive performance, they may be critical for some participants. 
The top issue features all have higher weight than the top identity features.  

\subsubsection{Synthesis: How do respondents' attitudes align with candidates positioning?}

Considering the first research question, both candidates mention identities more often than issues and both candidates mention neither in a majority of posts. Trump is particularly negative when talking about issues (i.e., attacking Harris' position, not defending his), and Harris is particularly negative in posts that include a mix of identity and issues.
Given the results in Table \ref{tab:results}, we might expect a candidate to mention issues more often. That is issues are much more predictive of party ID than identities, even when we model identity in a rich way by capturing multiple distinct dimensions (e.g., Group Association and Group Fate). 
However, Trump not only mentions identities more often, most of the time he mentions neither, and instead attacks the opposition candidate. 
This suggests that although issues are more predictive, neither candidate seems to think that speaking about the issues more than identities is a good strategy.
It is an open question as to whether candidates should spend more time discussing issues. 

Considering the second and third research questions, we ask if candidates speak about the groups and issues that are important to survey respondents. 
We see that the identity types Harris speaks about the most are gender, race, and age.  While some sub-groups are very Democratic, this does not align with the identities which are most predictive of party ID. Trump, in contrast, speaks the most about ``Americans'', then gender, then religious identities. This does seem to better align with identities which are predictive of being Republican, namely being born in the US and Protestant.  While identities were less predictive than issues, they still predict party ID reasonably well, and it appears that Trump had more focused messaging around key identities. 

When we look at Figure \ref{fig:issues} and Figure \ref{fig:issues_categories}, we see that the candidates focused on issues that party identifiers largely put value on. 
In Figure \ref{fig:issues}, we see that Harris spoke the most often about access to birth control and reproductive rights, followed by economic policy. Trump spoke about immigration and economic policy.
In Figure \ref{fig:issues_categories}, we can see that Democrats do care about abortion and maternity care and Republicans do care about issues related to economic policy, such as taxes and infrastructure. 
Interestingly, attitudes towards immigrants are highly predictive of party ID (as seen in Figure \ref{fig:features}), however Republicans don't put a lot of value on Immigration as an issue in Figure \ref{fig:qv} (although here is a case where the campaign may have affected what people care about, rather than the campaign focusing on what people ex-ante cared about). 

The candidates seem generally informed about the identities and issues that are important to respondents, and perhaps slightly less so to what is most predictive - or what differentiates Democrats from Republicans. 
 Harris, for example, mentioned reproductive rights. While this is important to Democrats and Republicans care about it less, the average issue position is not that different between Democrats and Republicans, especially on the milder interpretation of reproductive rights. That is, Harris spoke about a policy that may have enjoyed broad support, rather than an issue that only Democrats support, such as climate change and spending for renewable energy. 
 In contrast, Trump spoke about immigration, an issue which differentiates Democrats from Republicans both in terms of importance and position. 
 Both candidates spoke about economic policy; the difference is in the types of policies they discussed. 

 Thus, the candidates appear to be following different strategies. In short, our analysis suggests that (when not issuing attacks) Trump spoke to Republicans, emphasizing issues that are important to them and which they take significantly different positions on than Democrats, Harris spoke to a wider audience, mentioning issues which are important to Democrats but which they have a less distinctive stance on. Trump mentioned identities which are both common among and important to Republicans, and which differentiate them from Democrats. Harris mentioned identities which are important to Democrats, but not necessarily those which differentiate them from Republicans.  

\section{Discussion}
Quantitative descriptive data has a way of challenging accepted beliefs. Many media and political elite complain that Harris spent too much time talking about the rights of non-binary or transgender people\footnote{Bill Maher (\url{https://nypost.com/2024/11/16/us-news/bill-maher-says-r-d-democrats-blew-2024-election-with-woke-issues/}, James Carville (\url{https://www.newsweek.com/james-carville-harris-loss-woke-politics-1982035}), and Matt Yglesias (\url{https://www.slowboring.com/p/common-sense-manifesto-4-identity})}, but the data shows us that Harris avoided talking about that identity group: it was Trump attacking Harris for ties to them, not her claiming them, that was a salient part of the 2024 election cycle. This is highlighted in \figref{fig:smids}. Actually, the two candidates spent a similar amount of time on ``identity'' politics (\figref{fig:basics}) undermining this core punditry in 2025 regarding 2024.

Beyond the value of the core description of what the candidates were posting in regard to issues, identity, or neither, we also contribute new knowledge through our particular taxonomy of the interaction of these with negative affect. 
For example, we find that both candidates were more disparaging of the other candidate 
when discussing issues, than particular social group identities. 
This highlights that attack ads are not simply about the other candidate as a person, 
but about the policies they would implement and the ways in which they would support (or not) 
certain groups. While existing work has differentiated between negative ads focused on personality traits 
vs policy positions, with varying findings as to the relative effectiveness of each \citep{fridkin2008dimensions}, 
to the best of our knowledge no one has also explored how identity politics interacts with these 
two dimensions. This is an oversight, as we find that candidates are more likely to mention 
social identities than issues. Thus, future work should investigate the relative effectiveness 
of negative messaging when it frames one candidate as unable to further the interests 
of a particular group, without referencing particular policies. 

Some have suggested that it may be that no candidate offers what voters want \citep{fazekas2024presidential}. 
 Our survey results help provide more insight into this claim by showing what respondents care about and how they stand on a range of issues. We see that issues are much more predictive of party ID than identities, yet candidates speak about identities more (and neither issues or identities the most). 
This suggests there may be more room for candidates to speak about the issues that matter to potential voters. 

While we found that issues predicted party ID well, this may not straightforwardly support the idealized 
model of an informed voter with strong preferences. Instead, voters 
may have strong party identification, and work backwards from there to issues \citep{carsey2006changing}, 
as one's party has been shown to be a strong part of their social identity in some cases
 \citep{mason2018one,huddy2017political}. Yet, the extent to which partisan identification is an artifact of social identity is muddied by partisan identity itself; 
that is, the focus on partisan identity obscures other social group affiliations which may 
influence political behavior \citep{bernstein2005identity,hobsbawm1996identity}. 
For example  white identity, male identity, working class identity, and national identity dynamics have all been employed as an explanation of voting for Trump 
in the United States \citep{dignam2019misogynistic,buyuker2021race}, while similar constructs 
have been used to describe votes for Brexit \citep{bhambra2017brexit}.
Rural identity is widely associated with the Republican party in the United States, while urban or racial minority or female identity is associated with Democrats \citep{fudge2021place,rodden2019cities}.
While partisanship may activate different social identities, 
social identities can also drive allegiances to different parties \citep{pinsof2023strange}. 

To investigate the claim that issue positions are dictated by party ID and not the other way around, we fit a regression where we predicted support for each issue conditioned on identity and party ID (see \secref{identitytoissues}, entitled To what extent do social identities drive issue positions and importance?). 
Here, we find that even when controlling for party ID some identities 
are still predictive of issue position and importance. 
Together, these findings suggest that issue positions and importance
are not purely dictated by one's partisanship, 
but are at least partly influenced by one's actual lived experiences (a claim with implications outside of politics) \citep{mcconnell2018economic,druckman2021affective}. 
While party ID may be a social identity in its own right, 
our findings provide evidence that it is not the only identity, 
nor the sole determinant of one's policy preferences. 
This viewpoint has been partially advanced by \citet{castle2021partisanship}, 
who propose that when it is possible for voters to heuristically associate 
issues with identities they will do so.
This view is also complementary to Alliance Theory, 
which proposes that political belief systems are best explained 
as the results of alliances and rivalries to different groups \citep{pinsof2023strange}. 
Here, we see that issue position and importance is correlated with group identities.
Thus, rather than choosing a party based on alliances, one may chose a party based on its stated issue 
positions, where one assumes issue positions according to their social alliances. That is, successful models of political behavior will likely incorporate both expressive and instrumental aspects \citep{americanvoter}, and candidates employ both frames. 

\subsection{Limitations}
While focusing on elites, we do not investigate the political content posted on TikTok 
by everyday users. 
However, the user-driven content of social media messages
can have interesting political implications \citep{groshek2013public}, 
and as such is likely a fruitful direction for  future work. 

We focus on the most popular posts by either campaign. 
Our findings might differ if we were to consider all content. 
While we think it is valid to focus on the types of content 
an average apolitical viewer would encounter on TikTok, 
future work could focus on the types of content a more engaged viewer might see.

There is evidence that candidates may use different social media platforms 
differently \citep{rossini2017disentangling}. 
Thus, other work could compare and contrast strategies on TikTok 
to those found on other platforms. 

Our focus was primarily on how each team focused on either identities or issues. 
This distinction speaks to an ongoing debate between instrumental and expressive 
views of politics \citep{kalin2018think,fiorina1976voting}, with recent publications lending strength to both sides (instrumental \citep{orr2023affective} and expressive \citep{dias2022nature}). 
However, there are many other interesting categorizations one could make with this data. 
For example, one other dimension to explore would be the authenticity expressed by either candidate
\citep{enli2017twitter}. 
Finally, others have argued that partisanship is a stronger identity than others \citep{west2022partisanship,mason2018one}.
Our work instead engages with the question of how potential voters of each party express different identities and issue positions, 
without causally estimating whether party ID drives these expressions or the other way around. 

In our predictive model, we treat issues as the same. That is, we do not control for the fact that there may have been differences in how participants interpreted each issue question, and ignores other differences 
\citep{alvarez2004party,lachat2018party,carmines1980two}.
%and which between distinguishes ``easy'' and ``hard'' to understand issues \citep{carmines1980two}.
Future work may incorporate more complicated models of how issues interact with information about the parties themselves. Here, we refrained from potentially endogenous models.

In order to keep the predictive models interpretable, we opted for Logistic Regresssion. 
Others may find different results with more complex models. 
However, we did also explore different models, such as Random Forests, 
and found that we obtained similar and often better performance with Logistic Regression. 

\section{Additional methods background and details}
\label{sec:extra}
In order to model identity fully, we incorporate a range of related work. 
Similarly, we ask about a comprehensive set of policy issues. 
We share additional details about the \modelone{} and \modeltwo{} models here. 

\subsubsection{\modelone}
\modelone{}  allows us to explore the hypothesis 
that party identification is driven by belonging to other social groups. 
Here, we utilize a rich measurement of identity, which we term \textsc{Identity +}. 
This measurement utilizes several different dimensions of identity: 
Group Membership, Group Affiliation (i.e., which identities a person affiliates themselves), Group Fate (i.e,. which identities a person sees their fate linked to \citep{dawson1995behind}), Group Standing (which measures a sense of nostalgic deprivation \citep{gest2018roots}), and Group Support (which measures the extent to which a respondent would provide financial resources to different groups). These have been found to be important and non-overlapping measures of identity \citep{mcclain2009group}.

 A common starting point in measuring identity is \textbf{\textit{Group Membership}} (i.e., what group, or groups, a person assigns themselves to within all demographic categories). \citep{mcclain2009group} added the measure of perceived membership, which we term \textbf{\textit{Group Association}} (i.e., which groups a person associates themselves with). 
 In explaining Black politics in the United States, \citet{dawson1995behind} originated the term \textit{linked fate}. Linked fate describes the extent to which group members see their fate as linked to the fate of the group. He argues that even as their economic conditions change, 
Black voters see their fate linked to the fate of Black people in the United States. Linked fate, which we call \textbf{\textit{Group Fate}} has remained an essential measure of racial identity in politics \citep{simien2005race}.

Group Association and Fate, largely developed for Black voters, have been questioned for Hispanic and Asian voters \citep{junn2008asian,gershon2019intersectional}. However, that is not to say that race does not factor into politics for these groups  \citep{sadhwani2022influence,cox2019most}. 
For Asian and Hispanic voters the groups which they choose to ignore may be just as significant as those they choose to incorporate into their identities \citep{basler2008white,junn2008asian}. Recently, Hickel et al. found that Hispanic voters who identified as American, rather than Hispanic, were more likely to vote Republican \citep{hickel2020role}.

Simultaneously, scholars have called for a new understanding of white political identity \citep{jardina2019white}. Unlike previous work which has focused on out-group animus, Jardina has proposed that white identity is also explained by in-group attitudes, that is, those with higher white identity value being white, and have positive attitudes towards other whites. 
Additionally, recent work has investigated the extent to which some whites may see their group as a dispossessed minority \citep{jardina2021group}, suggesting the importance of group consciousness for this group as well \citep{berry2019white}.

The rise of the narrative of oppression in  political communication, along with the associated supports for populism in recent elections, 
bring up questions around the extent to which perceived societal standing 
%(which we will document as \textit{\textbf{Group Standing}}) 
influences political identity and behavior \citep{mondon2019whiteness,berlet2019rural,kazin2016trump}. 
Gest et al. measured subjective social, political, and economic status to be able to predict support for the Radical Right in the United States and United Kingdom \citep{gest2018roots}.  They found that nostalgic deprivation, a construct tightly related to ideas of a dispossessed minority, predicts support for the Radical Right. Whereas measures of societal standing are not generally included in models of partisanship, we incorporate this in a measure we term \textit{\textbf{Group Standing}}.

Individuals are influenced by a range of potentially conflicting social group identities \citep{brader2014cross,bergersen2018intersectionality,smith2007social,matos2023politics,bedolla2007intersections}. 
To model the fact that identities can compete and that voters are influenced by a range of identities, where respondents care about how politicians benefit certain groups \citep{americanvoter}, we introduce a measure of \textit{\textbf{Group Support}}. This measure captures which identities may be correlated with political action, as explained below. 

Incorporating this existing scholarship, \textsc{Identity+} is drawn from five categories of variables: \textbf{group membership}, 
\textbf{group association}, \textbf{group fate}, \textbf{group standing}, and \textbf{group support}:

\begin{itemize}
    \item \textbf{Group Membership} We start the survey by asking participants to select subgroup membership within 12 categories (and party identification):  age, gender, race and ethnicity, education, income,  religion, LGBTQ status, immigration status, union status, veteran status, place of residence, industry, and political party. %Next, for each of these identities (save political party) we ask a series of questions. 
    \item \textbf{Group Association}
To determine which groups the respondent most identifies with, we ask, ``These descriptions matter the MOST to me'' and ``These descriptions matter the LEAST to me''. The question is structured such that a respondent must select three of the possible subgroups that matter the most and three which matter the least. 
Each of the possible responses are generated  given the respondent's previous answers.  For example, a woman who wrote in her race as Salvadoran,\footnote{For this and the other identity strength items, if respondents indicated belonging to multiple groups we asked them to indicate which one they prefer that we refer to for the rest of the survey. So if someone indicated that they were both 'Black' and 'Vietnamese' we would ask them to indicate which group they prefer we use for the rest of the survey to refer to their race. We view this as a modest improvement over options which either force people to select one race, or refer to race in abstract terms. However, in future work we will refine the delivery of the survey to allow people to better describe multiple memberships. } with a college education, who works in the health industry, is a practicing protestant, and a union member, could say that the three most or least important descriptions to her are: ``I am Salvadoran'', ``I am a union cardholder'' and  ``My college degree''. This departs from surveys which allow respondents to select groups without identifying subgroups, for example, by selecting race/ethnicity, union status, or education as being important to them.

\item \textbf{Group Fate} Here, we follow a similar approach as for group association. For each group that a participant belongs to they must choose one of the following: “What happens generally to THIS GROUP in this country will have something to do with what happens in my life” and “What happens to this group will NOT impact my life; What happens to this group will impact my life A LITTLE; What happens to this group will impact my life SOMEWHAT; What happens to this group will impact my life A LOT”. We also constrain responses such that for at least one group and no more than three, they must select the most extreme options (NOT and A LOT). 

\item \textbf{Group Standing} To determine how one sees the groups within a societal context, we ask respondents to ``Place each group into categories indicating how important you think they are in American society.'' They are shown a diagram of concentric circles, and told that `` `1’ represents those that are considered the most central and important to society, whereas `4’ represents those that are considered the least central and important to society.''
This question was adapted from the question developed by \citet{gest2018roots}. In the original conception, people are asked to indicate the relative importance of large demographic groups, such as ``The Elderly''. Here, we are using standing as another dimension of identity, and asking respondents to indicate the relative importance only of those groups which they have indicated belonging to. 

\item \textbf{Group Support}
Finally, we ask participants to indicate how much they would favor different groups they belong to with financial support: both with a donation by a charity and spending by the government.
%Our hypothesis is that financial support will be another indication of how different groups factor into one's political identity. Especially, if respondents view politics as a form of redistribution, they should prefer parties which will distribute the most resources to the groups which are most important to them. 
We ask two questions: ``A charity has allocated funds to target groups of students to send to college and you've been asked to decide who should receive those funds among the following groups. Select the three groups you would target to send to college'' and ``The government has allocated funds to target groups of students to send to college and you've been asked to decide who should receive those funds among the following groups. Select the three groups you would target to send to college:''.
Now, they are shown the same groups as in previous questions, with the groups rephrased to be plural, for example, if a respondent selected \textit{Protestant} earlier, they would be shown the option to give to \textit{People of Protestant Faith}.
%Here, we are particularly concerned with how financial support can illuminate different aspects of identity. That is, rather than asking who people would prefer to give financial resources to across society, we constrain that they choose different groups amongst the groups that they belong to. 

\end{itemize}

\subsection{\modeltwo}
Here, we include positions on 34 issues across 14 political topics, 
from abortion to foreign policy. 
For each of the 34 issues covered in the second survey we ask four types of questions. The first asks respondents to select a position on an issue, 
the second asks respondents who they think gains and loses from an issue, 
the third asks their opinion about their perceptions of the issue positions of elites from the Democrat and Republican parties, 
and the fourth asks respondents to indicate how important 14 political topics are to them. 
Here, we analyze how respondents answer the issue position questions and the issue importance questions, questions two and three are the subject of future work. For example, an issue position question about abortion is as follows, \textit{Access to unconditional abortion before 15 weeks: a) Should BE protected by law, b) Should NOT be protected by law, or c) Don't know}.

In the issue importance question we group the 34 issues into 14 issue areas (in order to make the choices tractable for respondents). Respondents are presented with the following question. 
\textit{Imagine you have 20 \$100K tokens to allocate to lobby on behalf of your position in the following public policy categories, how would you allocate those 20 tokens among these public policy categories? \textbf{You must allocate all 20 tokens, but the max you can allocate for any given category is 10.}} The 14 issue areas are: 
Trade, Immigration, Healthcare, Voting Rights and Integrity, Infrastructure, Regulations, Foreign Relations, 
Taxes, Criminal Justice, Abortion and Maternity Care, Income inequality, Social Services for the Poor, Gun Rights and Safety, 
and Environmental Protection.

\bibliography{main}

\begin{thebibliography}{}

\bibitem[Achen, 1992]{achen1992social}
Achen, C.~H. (1992).
\newblock Social psychology, demographic variables, and linear regression: Breaking the iron triangle in voting research.
\newblock {\em Political behavior}, 14:195--211.

\bibitem[{Ad Fontes Media}, 2025]{adfontesmedia}
{Ad Fontes Media} (2025).
\newblock Ad fontes media.
\newblock Accessed: 2025-04-23.

\bibitem[Ahler and Sood, 2018]{ahler2018parties}
Ahler, D.~J. and Sood, G. (2018).
\newblock The parties in our heads: Misperceptions about party composition and their consequences.
\newblock {\em The Journal of Politics}, 80(3):964--981.

\bibitem[Allcott et~al., 2025]{allcott2025effects}
Allcott, H., Gentzkow, M., Levy, R., Crespo-Tenorio, A., Dumas, N., Mason, W., Moehler, D., Barbera, P., Brown, T.~W., Cisneros, J.~C., et~al. (2025).
\newblock The effects of political advertising on facebook and instagram before the 2020 us election.
\newblock Technical report, National Bureau of Economic Research.

\bibitem[Alvarez and Nagler, 2004]{alvarez2004party}
Alvarez, R.~M. and Nagler, J. (2004).
\newblock Party system compactness: Measurement and consequences.
\newblock {\em Political Analysis}, 12(1):46--62.

\bibitem[Ansolabehere and Iyengar, 1994]{ansolabehere1995going}
Ansolabehere, S. and Iyengar, S. (1994).
\newblock {\em Going negative: How attack ads shrink and polarize the electorate}.
\newblock Free Press, New York, NY.

\bibitem[Auter and Fine, 2016]{auter2016negative}
Auter, Z.~J. and Fine, J.~A. (2016).
\newblock Negative campaigning in the social media age: Attack advertising on facebook.
\newblock {\em Political Behavior}, 38(4):999--1020.

\bibitem[Bartels, 2002]{bartels2002beyond}
Bartels, L.~M. (2002).
\newblock Beyond the running tally: Partisan bias in political perceptions.
\newblock {\em Political behavior}, 24:117--150.

\bibitem[Basler, 2008]{basler2008white}
Basler, C. (2008).
\newblock White dreams and red votes: Mexican americans and the lure of inclusion in the republican party.
\newblock {\em Ethnic and Racial Studies}, 31(1):123--166.

\bibitem[Bedolla, 2007]{bedolla2007intersections}
Bedolla, L.~G. (2007).
\newblock Intersections of inequality: Understanding marginalization and privilege in the post-civil rights era.
\newblock {\em Politics \& Gender}, 3(2):232--248.

\bibitem[Bergersen et~al., 2018]{bergersen2018intersectionality}
Bergersen, M., Klar, S., and Schmitt, E. (2018).
\newblock Intersectionality and engagement among the lgbtq+ community.
\newblock {\em Journal of Women, Politics \& Policy}, 39(2):196--219.

\bibitem[Berlet and Sunshine, 2019]{berlet2019rural}
Berlet, C. and Sunshine, S. (2019).
\newblock Rural rage: the roots of right-wing populism in the united states.
\newblock {\em The Journal of Peasant Studies}, 46(3):480--513.

\bibitem[Bernstein, 2005]{bernstein2005identity}
Bernstein, M. (2005).
\newblock Identity politics.
\newblock {\em Annu. Rev. Sociol.}, 31:47--74.

\bibitem[Berry et~al., 2019]{berry2019white}
Berry, J.~A., Ebner, D., and Cornelius, M. (2019).
\newblock White identity politics: linked fate and political participation.
\newblock {\em Politics, Groups, and Identities}, 9(3):519--537.

\bibitem[Bhambra, 2017]{bhambra2017brexit}
Bhambra, G.~K. (2017).
\newblock Brexit, trump, and ‘methodological whiteness’: On the misrecognition of race and class.
\newblock {\em The British journal of sociology}, 68:S214--S232.

\bibitem[Bode et~al., 2020]{bode2020words}
Bode, L., Budak, C., and Ladd, J.~M. (2020).
\newblock {\em Words that matter: How the news and social media shaped the 2016 Presidential campaign}.
\newblock Brookings Institution Press.

\bibitem[Brader, 2005]{brader2005striking}
Brader, T. (2005).
\newblock Striking a responsive chord: How political ads motivate and persuade voters by appealing to emotions.
\newblock {\em American journal of political science}, 49(2):388--405.

\bibitem[Brader et~al., 2014]{brader2014cross}
Brader, T., Tucker, J.~A., and Therriault, A. (2014).
\newblock Cross pressure scores: An individual-level measure of cumulative partisan pressures arising from social group memberships.
\newblock {\em Political Behavior}, 36(1):23--51.

\bibitem[Bronstein, 2013]{bronstein2013like}
Bronstein, J. (2013).
\newblock Like me! analyzing the 2012 presidential candidates’ facebook pages.
\newblock {\em Online Information Review}, 37(2):173--192.

\bibitem[Brown and Enos, 2021]{brown2021measurement}
Brown, J.~R. and Enos, R.~D. (2021).
\newblock The measurement of partisan sorting for 180 million voters.
\newblock {\em Nature Human Behaviour}, 5(8):998--1008.

\bibitem[Buccoliero et~al., 2020]{buccoliero2020twitter}
Buccoliero, L., Bellio, E., Crestini, G., and Arkoudas, A. (2020).
\newblock Twitter and politics: Evidence from the us presidential elections 2016.
\newblock {\em Journal of Marketing Communications}, 26(1):88--114.

\bibitem[Buyuker et~al., 2021]{buyuker2021race}
Buyuker, B., D'Urso, A.~J., Filindra, A., and Kaplan, N.~J. (2021).
\newblock Race politics research and the american presidency: thinking about white attitudes, identities and vote choice in the trump era and beyond.
\newblock {\em Journal of Race, Ethnicity, and Politics}, 6(3):600--641.

\bibitem[Campbell et~al., 1960]{americanvoter}
Campbell, A., Converse, P.~E., Miller, W.~E., and Stokes, D.~E. (1960).
\newblock {\em The American Voter}.
\newblock University of Chicago Press.

\bibitem[Campbell, 2018]{campbell2018polarized}
Campbell, J.~E. (2018).
\newblock {\em Polarized: Making sense of a divided America}.
\newblock Princeton University Press.

\bibitem[Carmines and Stimson, 1980]{carmines1980two}
Carmines, E.~G. and Stimson, J.~A. (1980).
\newblock The two faces of issue voting.
\newblock {\em American Political Science Review}, 74(1):78--91.

\bibitem[Carsey and Layman, 2006]{carsey2006changing}
Carsey, T.~M. and Layman, G.~C. (2006).
\newblock Changing sides or changing minds? {P}arty identification and policy preferences in the american electorate.
\newblock {\em American Journal of Political Science}, 50(2):464--477.

\bibitem[Castle and Stepp, 2021]{castle2021partisanship}
Castle, J.~J. and Stepp, K.~K. (2021).
\newblock Partisanship, religion, and issue polarization in the united states: A reassessment.
\newblock {\em Political Behavior}, pages 1--25.

\bibitem[Clinton and Lapinski, 2004]{clinton2004targeted}
Clinton, J.~D. and Lapinski, J.~S. (2004).
\newblock “targeted” advertising and voter turnout: An experimental study of the 2000 presidential election.
\newblock {\em The Journal of Politics}, 66(1):69--96.

\bibitem[Congleton, 1991]{congleton1991information}
Congleton, R.~D. (1991).
\newblock Information, special interests, and single-issue voting.
\newblock {\em Public Choice}, 69(1):39--49.

\bibitem[Cox, 2019]{cox2019most}
Cox, K. (2019).
\newblock Most us adults feel what happens to their own racial or ethnic group affects them personally.

\bibitem[Davis and Mason, 2016]{davis2016sorting}
Davis, N.~T. and Mason, L. (2016).
\newblock Sorting and the split-ticket: Evidence from presidential and subpresidential elections.
\newblock {\em Political Behavior}, 38:337--354.

\bibitem[Dawson, 1995]{dawson1995behind}
Dawson, M.~C. (1995).
\newblock {\em Behind the Mule: Race and Class in African-American Politics}.
\newblock Princeton University Press.

\bibitem[de~Boer et~al., 2012]{de2012social}
de~Boer, N., S{\"u}tfeld, H., and Groshek, J. (2012).
\newblock Social media and personal attacks: A comparative perspective on co-creation and political advertising in presidential campaigns on youtube.
\newblock {\em First Monday}.

\bibitem[Dias and Lelkes, 2022]{dias2022nature}
Dias, N. and Lelkes, Y. (2022).
\newblock The nature of affective polarization: Disentangling policy disagreement from partisan identity.
\newblock {\em American Journal of Political Science}, 66(3):775--790.

\bibitem[Dignam and Rohlinger, 2019]{dignam2019misogynistic}
Dignam, P.~A. and Rohlinger, D.~A. (2019).
\newblock Misogynistic men online: How the red pill helped elect trump.
\newblock {\em Signs: Journal of Women in Culture and Society}, 44(3):589--612.

\bibitem[Donald et~al., 2002]{donald2002partisan}
Donald, G., Bradley, P., and Eric, S. (2002).
\newblock Partisan hearts and minds.

\bibitem[Druckman et~al., 2021]{druckman2021affective}
Druckman, J.~N., Klar, S., Krupnikov, Y., Levendusky, M., and Ryan, J.~B. (2021).
\newblock How affective polarization shapes americans’ political beliefs: A study of response to the covid-19 pandemic.
\newblock {\em Journal of Experimental Political Science}, 8(3):223--234.

\bibitem[Enli, 2017]{enli2017twitter}
Enli, G. (2017).
\newblock Twitter as arena for the authentic outsider: Exploring the social media campaigns of trump and clinton in the 2016 us presidential election.
\newblock {\em European journal of communication}, 32(1):50--61.

\bibitem[Eshbaugh-Soha, 2015]{eshbaugh2015traditional}
Eshbaugh-Soha, M. (2015).
\newblock Traditional media, social media, and different presidential campaign messages.
\newblock {\em Controlling the message: New media in American political campaigns}, 136.

\bibitem[Fazekas and Hatemi, 2024]{fazekas2024presidential}
Fazekas, Z. and Hatemi, P.~K. (2024).
\newblock Presidential candidates nobody wants?
\newblock {\em Presidential Studies Quarterly}, 54(1):115--129.

\bibitem[Fine and Hunt, 2023]{fine2023negativity}
Fine, J.~A. and Hunt, M.~F. (2023).
\newblock Negativity and elite message diffusion on social media.
\newblock {\em Political Behavior}, 45(3):955--973.

\bibitem[Fiorina, 1976]{fiorina1976voting}
Fiorina, M.~P. (1976).
\newblock The voting decision: instrumental and expressive aspects.
\newblock {\em The journal of politics}, 38(2):390--413.

\bibitem[Fournier et~al., 2003]{fournier2003issue}
Fournier, P., Blais, A., Nadeau, R., Gidengil, E., and Nevitte, N. (2003).
\newblock Issue importance and performance voting.
\newblock {\em Political Behavior}, 25:51--67.

\bibitem[Fowler et~al., 2021]{fowler2021political}
Fowler, E.~F., Franz, M.~M., Martin, G.~J., Peskowitz, Z., and Ridout, T.~N. (2021).
\newblock Political advertising online and offline.
\newblock {\em American Political Science Review}, 115(1):130--149.

\bibitem[Franz et~al., 2020]{franz2020issue}
Franz, M.~M., Franklin~Fowler, E., Ridout, T., and Wang, M.~Y. (2020).
\newblock The issue focus of online and television advertising in the 2016 presidential campaign.
\newblock {\em American Politics Research}, 48(1):175--196.

\bibitem[Freedman et~al., 2004]{freedman2004campaign}
Freedman, P., Franz, M., and Goldstein, K. (2004).
\newblock Campaign advertising and democratic citizenship.
\newblock {\em American Journal of Political Science}, 48(4):723--741.

\bibitem[Fridkin and Kenney, 2004]{fridkin2004negative}
Fridkin, K.~L. and Kenney, P.~J. (2004).
\newblock Do negative messages work? the impact of negativity on citizens’ evaluations of candidates.
\newblock {\em American politics research}, 32(5):570--605.

\bibitem[Fridkin and Kenney, 2008]{fridkin2008dimensions}
Fridkin, K.~L. and Kenney, P.~J. (2008).
\newblock The dimensions of negative messages.
\newblock {\em American Politics Research}, 36(5):694--723.

\bibitem[Fudge and Armaly, 2021]{fudge2021place}
Fudge, D. and Armaly, M.~T. (2021).
\newblock Place or partisanship? the conditional role of identities on candidate support.
\newblock {\em American Politics Research}, 49(6):655--665.

\bibitem[Fukuyama, 2018]{fukuyama2018against}
Fukuyama, F. (2018).
\newblock Against identity politics: The new tribalism and the crisis of democracy.
\newblock {\em Foreign Aff.}, 97:90.

\bibitem[Fulgoni et~al., 2016]{fulgoni2016power}
Fulgoni, G.~M., Lipsman, A., and Davidsen, C. (2016).
\newblock The power of political advertising: Lessons for practitioners: How data analytics, social media, and creative strategies shape us presidential election campaigns.
\newblock {\em Journal of Advertising Research}, 56(3):239--244.

\bibitem[Galasso et~al., 2023]{galasso2023positive}
Galasso, V., Nannicini, T., and Nunnari, S. (2023).
\newblock Positive spillovers from negative campaigning.
\newblock {\em American Journal of Political Science}, 67(1):5--21.

\bibitem[Geer, 2012]{geer2012news}
Geer, J.~G. (2012).
\newblock The news media and the rise of negativity in presidential campaigns.
\newblock {\em PS: Political Science \& Politics}, 45(3):422--427.

\bibitem[Gershon et~al., 2019]{gershon2019intersectional}
Gershon, S.~A., Montoya, C., Bejarano, C., and Brown, N. (2019).
\newblock Intersectional linked fate and political representation.
\newblock {\em Politics, Groups, and Identities}, 7(3):642--653.

\bibitem[Gest et~al., 2018]{gest2018roots}
Gest, J., Reny, T., and Mayer, J. (2018).
\newblock Roots of the radical right: Nostalgic deprivation in the united states and britain.
\newblock {\em Comparative Political Studies}, 51(13):1694--1719.

\bibitem[Gil~de Z{\'u}{\~n}iga et~al., 2017]{gil2017effects}
Gil~de Z{\'u}{\~n}iga, H., Weeks, B., and Ard{\`e}vol-Abreu, A. (2017).
\newblock Effects of the news-finds-me perception in communication: Social media use implications for news seeking and learning about politics.
\newblock {\em Journal of computer-mediated communication}, 22(3):105--123.

\bibitem[Glasgow and Alvarez, 2000]{glasgow2000uncertainty}
Glasgow, G. and Alvarez, R.~M. (2000).
\newblock Uncertainty and candidate personality traits.
\newblock {\em American Politics Quarterly}, 28(1):26--49.

\bibitem[Gleibs et~al., 2018]{gleibs2018identity}
Gleibs, I.~H., Hendricks, K., and Kurz, T. (2018).
\newblock Identity mediators: Leadership and identity construction in campaign speeches of american presidential candidates' spouses.
\newblock {\em Political Psychology}, 39(4):939--956.

\bibitem[Groshek and Al-Rawi, 2013]{groshek2013public}
Groshek, J. and Al-Rawi, A. (2013).
\newblock Public sentiment and critical framing in social media content during the 2012 us presidential campaign.
\newblock {\em Social Science Computer Review}, 31(5):563--576.

\bibitem[Grossmann and Hopkins, 2015]{grossmann2015ideological}
Grossmann, M. and Hopkins, D.~A. (2015).
\newblock Ideological republicans and group interest democrats: The asymmetry of american party politics.
\newblock {\em Perspectives on Politics}, 13(1):119--139.

\bibitem[Haenschen, 2023]{haenschen2023conditional}
Haenschen, K. (2023).
\newblock The conditional effects of microtargeted facebook advertisements on voter turnout.
\newblock {\em Political Behavior}, 45(4):1661--1681.

\bibitem[Hersh, 2015]{hersh2015hacking}
Hersh, E.~D. (2015).
\newblock {\em Hacking the electorate: How campaigns perceive voters}.
\newblock Cambridge University Press.

\bibitem[Hickel~Jr et~al., 2020]{hickel2020role}
Hickel~Jr, F.~R., Alamillo, R., Oskooii, K.~A., and Collingwood, L. (2020).
\newblock The role of identity prioritization: Why some latinx support restrictionist immigration policies and candidates.
\newblock {\em Public Opinion Quarterly}, 84(4):860--891.

\bibitem[Highton and Kam, 2011]{highton2011long}
Highton, B. and Kam, C.~D. (2011).
\newblock The long-term dynamics of partisanship and issue orientations.
\newblock {\em The Journal of Politics}, 73(1):202--215.

\bibitem[Hindarto, 2022]{hindarto2022tiktok}
Hindarto, I.~H. (2022).
\newblock Tiktok and political communication of youth: a systematic review.
\newblock {\em JRP (Jurnal Review Politik)}, 12(2):146--176.

\bibitem[Hobsbawm, 1996]{hobsbawm1996identity}
Hobsbawm, E. (1996).
\newblock Identity politics and the left.
\newblock {\em new left review}, pages 38--47.

\bibitem[Huddy and Bankert, 2017]{huddy2017political}
Huddy, L. and Bankert, A. (2017).
\newblock Political partisanship as a social identity.

\bibitem[Huddy et~al., 2015]{huddy2015expressive}
Huddy, L., Mason, L., and Aar{\o}e, L. (2015).
\newblock Expressive partisanship: Campaign involvement, political emotion, and partisan identity.
\newblock {\em American Political Science Review}, 109(1):1--17.

\bibitem[Iyengar et~al., 2019]{iyengar2019origins}
Iyengar, S., Lelkes, Y., Levendusky, M., Malhotra, N., and Westwood, S.~J. (2019).
\newblock The origins and consequences of affective polarization in the united states.
\newblock {\em Annual review of political science}, 22(1):129--146.

\bibitem[Iyengar and Westwood, 2015]{iyengar2015fear}
Iyengar, S. and Westwood, S.~J. (2015).
\newblock Fear and loathing across party lines: New evidence on group polarization.
\newblock {\em American journal of political science}, 59(3):690--707.

\bibitem[Jamieson et~al., 2000]{jamieson2000eliminate}
Jamieson, K.~H., Waldman, P., and Sherr, S. (2000).
\newblock Eliminate the negative? categories of analysis for political advertisements.
\newblock {\em Crowded airwaves: Campaign advertising in elections}, pages 44--64.

\bibitem[Jardina, 2019]{jardina2019white}
Jardina, A. (2019).
\newblock {\em White identity politics}.
\newblock Cambridge University Press.

\bibitem[Jardina, 2021]{jardina2021group}
Jardina, A. (2021).
\newblock In-group love and out-group hate: White racial attitudes in contemporary us elections.
\newblock {\em Political Behavior}, 43(4):1535--1559.

\bibitem[Jordan et~al., 2018]{jordan20182016}
Jordan, K.~N., Pennebaker, J.~W., and Ehrig, C. (2018).
\newblock The 2016 us presidential candidates and how people tweeted about them.
\newblock {\em Sage Open}, 8(3):2158244018791218.

\bibitem[Junn and Masuoka, 2008]{junn2008asian}
Junn, J. and Masuoka, N. (2008).
\newblock Asian american identity: Shared racial status and political context.
\newblock {\em Perspectives on Politics}, 6(4):729--740.

\bibitem[Kalin and Sambanis, 2018]{kalin2018think}
Kalin, M. and Sambanis, N. (2018).
\newblock How to think about social identity.
\newblock {\em Annual Review of Political Science}, 21:239--257.

\bibitem[Kane et~al., 2021]{kane2021s}
Kane, J.~V., Mason, L., and Wronski, J. (2021).
\newblock Who’s at the party? group sentiments, knowledge, and partisan identity.
\newblock {\em The Journal of Politics}, 83(4):1783--1799.

\bibitem[Kaplan et~al., 2006]{kaplan2006dialogue}
Kaplan, N., Park, D.~K., and Ridout, T.~N. (2006).
\newblock Dialogue in american political campaigns? an examination of issue convergence in candidate television advertising.
\newblock {\em American Journal of Political Science}, 50(3):724--736.

\bibitem[Kazin, 2016]{kazin2016trump}
Kazin, M. (2016).
\newblock Trump and american populism: Old whine, new bottles.
\newblock {\em Foreign Affairs}, 95(6):17--24.

\bibitem[Kidder, 2016]{kidder2016college}
Kidder, J.~L. (2016).
\newblock College republicans and conservative social identity.
\newblock {\em Sociological Perspectives}, 59(1):177--200.

\bibitem[Klug et~al., 2021]{klug2021trick}
Klug, D., Qin, Y., Evans, M., and Kaufman, G. (2021).
\newblock Trick and please. a mixed-method study on user assumptions about the tiktok algorithm.
\newblock In {\em Proceedings of the 13th ACM web science conference 2021}, pages 84--92.

\bibitem[Lachat and Wagner, 2018]{lachat2018party}
Lachat, R. and Wagner, A. (2018).
\newblock How party characteristics drive voters' evaluation criteria.
\newblock {\em Electoral Studies}, 55:11--20.

\bibitem[Larsson et~al., 2024]{larsson2024calls}
Larsson, A.~O., T{\o}nnesen, H., Magin, M., and Skogerb{\o}, E. (2024).
\newblock Calls to (what kind of?) action: A framework for comparing political actors’ campaign strategies across social media platforms.
\newblock {\em new media \& society}, page 14614448241229156.

\bibitem[Latimer, 1984]{latimer1984policy}
Latimer, M.~K. (1984).
\newblock Policy issues and personal images in political advertising in a state election.
\newblock {\em Journalism Quarterly}, 61(4):776--852.

\bibitem[Lau et~al., 1999]{lau1999effects}
Lau, R.~R., Sigelman, L., Heldman, C., and Babbitt, P. (1999).
\newblock The effects of negative political advertisements: A meta-analytic assessment.
\newblock {\em American Political Science Review}, 93(4):851--875.

\bibitem[Lavigne, 2021]{lavigne2021strengthening}
Lavigne, M. (2021).
\newblock Strengthening ties: The influence of microtargeting on partisan attitudes and the vote.
\newblock {\em Party Politics}, 27(5):965--976.

\bibitem[Lelkes, 2016]{lelkes2016mass}
Lelkes, Y. (2016).
\newblock Mass polarization: Manifestations and measurements.
\newblock {\em Public Opinion Quarterly}, 80(S1):392--410.

\bibitem[Lenz~Gabriel, 2012]{lenz2012follow}
Lenz~Gabriel, S. (2012).
\newblock Follow the leader? how voters respond to politicians’ policies and performance.
\newblock {\em Chicago Studies in American Politics. Chicago, IL: University of Chicago Press}.

\bibitem[Lewis-Beck, 2008]{lewis2008american}
Lewis-Beck, M.~S. (2008).
\newblock {\em The American voter revisited}.
\newblock University of Michigan Press.

\bibitem[Martin and Nai, 2024]{martin2024deepening}
Martin, D. and Nai, A. (2024).
\newblock Deepening the rift: Negative campaigning fosters affective polarization in multiparty elections.
\newblock {\em Electoral Studies}, 87:102745.

\bibitem[Mason and Wronski, 2018]{mason2018one}
Mason, L. and Wronski, J. (2018).
\newblock One tribe to bind them all: How our social group attachments strengthen partisanship.
\newblock {\em Political Psychology}, 39:257--277.

\bibitem[Matos et~al., 2023]{matos2023politics}
Matos, Y., Greene, S., and Sanbonmatsu, K. (2023).
\newblock The politics of “women of color”: a group identity worth investigating.
\newblock {\em Politics, Groups, and Identities}, 11(3):549--570.

\bibitem[McClain et~al., 2009]{mcclain2009group}
McClain, P.~D., Johnson~Carew, J.~D., Walton~Jr, E., and Watts, C.~S. (2009).
\newblock Group membership, group identity, and group consciousness: Measures of racial identity in american politics?
\newblock {\em Annual Review of Political Science}, 12:471--485.

\bibitem[McConnell et~al., 2018]{mcconnell2018economic}
McConnell, C., Margalit, Y., Malhotra, N., and Levendusky, M. (2018).
\newblock The economic consequences of partisanship in a polarized era.
\newblock {\em American Journal of Political Science}, 62(1):5--18.

\bibitem[MClain et~al., 2024]{pew2024}
MClain, C., Anderson, M., and Galles-Watnick, R. (2024).
\newblock How tiktok users view, experience the platform.

\bibitem[Metzgar and Maruggi, 2009]{metzgar2009social}
Metzgar, E. and Maruggi, A. (2009).
\newblock Social media and the 2008 us presidential election.
\newblock {\em Journal of new communications research}, 4(1).

\bibitem[Mondon and Winter, 2019]{mondon2019whiteness}
Mondon, A. and Winter, A. (2019).
\newblock Whiteness, populism and the racialisation of the working class in the united kingdom and the united states.
\newblock {\em Identities}, 26(5):510--528.

\bibitem[Orr et~al., 2023]{orr2023affective}
Orr, L.~V., Fowler, A., and Huber, G.~A. (2023).
\newblock Is affective polarization driven by identity, loyalty, or substance?
\newblock {\em American Journal of Political Science}, 67(4):948--962.

\bibitem[Pinsof et~al., 2023]{pinsof2023strange}
Pinsof, D., Sears, D.~O., and Haselton, M.~G. (2023).
\newblock Strange bedfellows: The alliance theory of political belief systems.
\newblock {\em Psychological Inquiry}, 34(3):139--160.

\bibitem[Rodden, 2019]{rodden2019cities}
Rodden, J.~A. (2019).
\newblock {\em Why cities lose: The deep roots of the urban-rural political divide}.
\newblock Basic Books.

\bibitem[Rossini et~al., 2017]{rossini2017disentangling}
Rossini, P. G.~C., Hemsley, J., Tanupabrungsun, S., Zhang, F., Robinson, J., and Stromer-Galley, J. (2017).
\newblock Social media, u.s. presidential campaigns, and public opinion polls: Disentangling effects.
\newblock In {\em Proceedings of the 8th International Conference on Social Media \& Society}, SMSociety17, New York, NY, USA. Association for Computing Machinery.

\bibitem[Ruckelshaus, 2022]{ruckelshaus2022kind}
Ruckelshaus, J. (2022).
\newblock What kind of identity is partisan identity?“social” versus “political” partisanship in divided democracies.
\newblock {\em American Political Science Review}, 116(4):1477--1489.

\bibitem[Sadhwani, 2022]{sadhwani2022influence}
Sadhwani, S. (2022).
\newblock The influence of candidate race and ethnicity: the case of asian americans.
\newblock {\em Politics, Groups, and Identities}, 10(4):616--652.

\bibitem[Sandvig et~al., 2014]{sandvig2014auditing}
Sandvig, C., Hamilton, K., Karahalios, K., and Langbort, C. (2014).
\newblock Auditing algorithms: Research methods for detecting discrimination on internet platforms.
\newblock {\em Data and discrimination: converting critical concerns into productive inquiry}, 22(2014):4349--4357.

\bibitem[Savoy, 2018]{savoy2018analysis}
Savoy, J. (2018).
\newblock Analysis of the style and the rhetoric of the 2016 us presidential primaries.
\newblock {\em Digital Scholarship in the Humanities}, 33(1):143--159.

\bibitem[Shafranek, 2020]{shafranek2020political}
Shafranek, R.~M. (2020).
\newblock Political consequences of partisan prejudice.
\newblock {\em Political Psychology}, 41(1):35--51.

\bibitem[Sides and Vavreck, 2014]{sides2014gamble}
Sides, J. and Vavreck, L. (2014).
\newblock The gamble: Choice and chance in the 2012 presidential election-updated edition.
\newblock In {\em The Gamble}. Princeton University Press.

\bibitem[Simien, 2005]{simien2005race}
Simien, E.~M. (2005).
\newblock Race, gender, and linked fate.
\newblock {\em Journal of Black Studies}, 35(5):529--550.

\bibitem[Smith, 2007]{smith2007social}
Smith, T.~W. (2007).
\newblock Social identity and socio-demographic structure.
\newblock {\em International Journal of Public Opinion Research}, 19(3):380--390.

\bibitem[Stoetzer and Zittlau, 2020]{stoetzer2020vote}
Stoetzer, L.~F. and Zittlau, S. (2020).
\newblock Vote choice and the nonseparability of economic and social issues.
\newblock {\em Public Opinion Quarterly}, 84(1):158--170.

\bibitem[Stromer-Galley et~al., 2021]{stromer2021political}
Stromer-Galley, J., Rossini, P., Hemsley, J., Bolden, S.~E., and McKernan, B. (2021).
\newblock Political messaging over time: A comparison of us presidential candidate facebook posts and tweets in 2016 and 2020.
\newblock {\em Social Media+ Society}, 7(4):20563051211063465.

\bibitem[Van~Steenburg and Guzm{\'a}n, 2019]{van2019influence}
Van~Steenburg, E. and Guzm{\'a}n, F. (2019).
\newblock The influence of political candidate brands during the 2012 and 2016 us presidential elections.
\newblock {\em European Journal of Marketing}, 53(12):2629--2656.

\bibitem[West and Iyengar, 2022]{west2022partisanship}
West, E.~A. and Iyengar, S. (2022).
\newblock Partisanship as a social identity: Implications for polarization.
\newblock {\em Political Behavior}, 44(2):807--838.

\bibitem[Westwood and Lelkes, 2024]{westwood2024pulse}
Westwood, S. and Lelkes, Y. (2024).
\newblock America’s political pulse.
\newblock Accessed: 2025-08-05.

\bibitem[Wright~Jr and Berkman, 1986]{wright1986candidates}
Wright~Jr, G.~C. and Berkman, M.~B. (1986).
\newblock Candidates and policy in united states senate elections.
\newblock {\em American Political Science Review}, 80(2):567--588.

\bibitem[Wurst et~al., 2023]{wurst2023mobilization}
Wurst, A.-K., Pohl, K., and Ha{\ss}ler, J. (2023).
\newblock Mobilization in the context of campaign functions and citizen participation.
\newblock {\em Media and Communication}, 11(3):129--140.

\bibitem[Yu et~al., 2024]{yu2024partisanship}
Yu, X., Wojcieszak, M., and Casas, A. (2024).
\newblock Partisanship on social media: In-party love among american politicians, greater engagement with out-party hate among ordinary users.
\newblock {\em Political Behavior}, 46(2):799--824.

\end{thebibliography}

\begin{appendices}

\section{Sock puppets setup}

See Table \ref{tab:influencers} for lifestyle influencers we chose for accounts to follow.

\noindent See Table \ref{tab:news} for news channels we chose for accounts to follow.

\begin{table}[]
\noindent\adjustbox{max width=.79\columnwidth}{%
    \begin{adjustbox}{width=0.8\textwidth,center}{
\def\sym#1{\ifmmode^{#1}\else\(^{#1}\)\fi}
\begin{tabular}{lll}
\toprule
Category                                   & author\_name                    & author\_id               \\
\midrule
Sustainability           & HomesteadDonegal        & mirendarosenberg     \\
                                           & Ken Russell             & kenforflorida        \\
                                           & Alaina Wood             & thegarbagequeen      \\
                                           & Steeze365Daily          & steeze365daily       \\
                                           & thesorrygirls           & thesorrygirls        \\
                                           & ReLauren                & relauren             \\
                                           & Reallaurinda            & reallaurinda         \\
                                           & Ashley Diedenhofen      & sciencebyashley      \\
                                           & TheNotoriousKIA         & thenotoriouskia      \\
                                           & Phil Sustainability     & philsustainability   \\
                                           & Brennan Kai             & brennan.kai          \\ \midrule
Wellness              &    Jessamyn $|$ The Underbelly Yoga                     & mynameisjessamyn     \\
                                           & Mari Llewellyn                        & marillewellyn        \\
                                           &  Arielle Lorre                       & ariellelorre         \\
                                           &          Dr. Will Cole               & drwillcole           \\
                                           &     Micheline Maalouf Therapist                    & micheline.maalouf    \\
                                           &  Staci Tanouye, MD                       & dr.staci.t           \\
                                           & Andrew Huberman                        & hubermanlab          \\
                                           &  Fiona                       & feelgoodwith\_fi     \\
                                           &   Steph Grasso, MS, RD                      & stephgrassodietitian \\
                                           &    Daniel                     & mrduku               \\ \midrule
DIY / Home improvement   & Molly Miller            & therenegadehome      \\
                                           & Bong Bain               & wildheartshome       \\
                                           & Lilly                   & thefurnituredoctor   \\
                                           & Christine Higg          & \_forthehome         \\
                                           & Joanna Gaines           & joannagaines         \\
                                           & Lone Fox                & lonefoxhome          \\
                                           & Kylie Katich            & kyliekatich          \\
                                           &  CASSMAKESHOME $|$ HOME \& DIY                       & cassmakeshome        \\
                                           &  Renovating Our Home                       & renovatingourhome    \\
                                           &  Jay Munee DIY                       & jaymuneediy          \\
                                           &   kelsey                      & kelseydarragh        \\
                                           &  Abby                       & abby\_roadhome       \\
                                           &  Contractor Ken                       & contractorken        \\
                                           &  ReallyVeryCrunchy                       & reallyverycrunchy    \\
                                           &  THE FLIPPED PIECE                       & theflippedpiece      \\
                                           &  Jeff Thorman                       & homerenovisiondiy    \\
                                           &  Bro Builds                       & bro\_builds          \\ \midrule
Tech                   & koharotv                & koharotvreal         \\
                                           & Tyler Morgan            & hitomidocameraroll   \\
                                           & Jimena con jota         & soyjimenaconjota     \\
                                           & CHIP                    & chip\_de             \\
                                           & Mark's Tech             & markstech            \\
                                           & TheAsianJC              & theasianjc           \\
                                           & Lucas VRTech            & lucas\_vrtech        \\
                                           & Marques Brownlee        & mkbhd                \\
                                           & Unbox Therapy           & unboxtherapyofficial \\
                                           & Austin Evans            & austintechtips       \\
                                           & iJustine                & ijustine             \\
                                           & Kevin Stratvert         & kevinstratvert       \\
                                           & Sara Dietschy           & saradietschy         \\ \midrule
College Sports         & Olivia Dunne            & livvy                \\
                                           & Paige Beukers           & paigebueckers        \\
                                           & Hanna \& Haley Cavinder & cavindertwins        \\
                                           & Khoi Young              & khoiyoung7           \\
                                           & Frederick Richards      & frederickflips       \\
                                           & Shedeur Sanders         & shedeursanders       \\
                                           & Angel Reese             & angelreese10         \\
                                           & Caitlin Clark           & caitlin.clark22      \\
                                           & Bronny James            & bronny               \\
                                           & A.J. Henning            & ajhenning            \\ \midrule
Hunting/fishing/outdoors &     BlacktipH Fishing                     & blacktiph            \\
                                           &   Ryan Izquierdo                      & ryanizfishing        \\
                                           &   jetreef                      & jetreef              \\
                                           &   RAWW Fishing                      & rawwfishingyt        \\
                                           &   kickintheirbasstv                      & kickintheirbasstv    \\
                                           &  Outdoors Weekly                       & outdoorsweekly       \\
                                           &  Frederick Penney                       & frederick            \\
                                           &  Becky Granola Girl                      & bonjourbecky         \\
                                           &   Keith Paluso                      & thisiskeithpaluso    \\
                                           &    Kween werK                       & kweenwerk    \\ \bottomrule       
\end{tabular}
}
\end{adjustbox}
    }
    \caption{Lifestyle influencers we chose for accounts to follow. These were chosen as popular, and generally apolitical accounts as far as we could tell.}
    \label{tab:influencers}
\end{table}

\begin{table}[htb!]
    \centering
    \begin{tabular}{llrr}
\toprule
author\_name            & author\_id         & Reliability & Bias  \\
\midrule
New York Times  & nytimes        & 41.04       & -8.07 \\
NBC             & nbcnews        & 42.80        & -5.64 \\
Washington Post & washingtonpost & 38.83       & -6.93 \\
PBS News        & pbsnews        & 43.32       & -4.05 \\
ABC             & abcnews        & 44.80        & -3.00    \\
CBS             & cbsnews        & 42.03       & -2.72 \\
NPR             & npr            & 43.09       & -4.17 \\
BBC News        & bbcnews        & 44.73       & -1.35 \\
Yahoo News      & yahoonews      & 40.94       & -5.63 \\
USA Today       & usatoday       & 40.86       & -4.06 \\
\bottomrule
\end{tabular}
    \caption{News channels we chose for accounts to follow. Reliability scores and bias scores are extracted from the Media Bias Chart \citep{adfontesmedia}. Reliability scores for articles and shows are on a scale of 0-64. Scores above 40 are generally good. Bias scores for articles and shows are on a scale of -42 to +42, with higher negative scores being more left, higher positive scores being more right, and scores closer to zero being minimally biased, equally balanced, or exhibiting a centrist bias. Typically, a publication would be considered centrist if the score is between -10 and +10, left-oriented if the score is -10 or less and right-leaning if the score is +10 or more.}
    \label{tab:news}
\end{table}

\section{Links to survey documents}\label{sec:survey_doc}

See the full identity survey  document \href{https://drive.google.com/file/d/10GZUlG9EeMFcJu3vM7i8XD-GD2xFmn6w/view?usp=drive_link}{here}.

\noindent See the full issues survey document \href{https://drive.google.com/file/d/10GZUlG9EeMFcJu3vM7i8XD-GD2xFmn6w/view?usp=drive_link}{here}.

\newpage
\section{Additional analyses}\label{sec:add_analysis}

\subsection{Robustness Check}
\label{sec:robustness}
\begin{figure}[hb!]
    \centering
    \includegraphics[width=\linewidth]{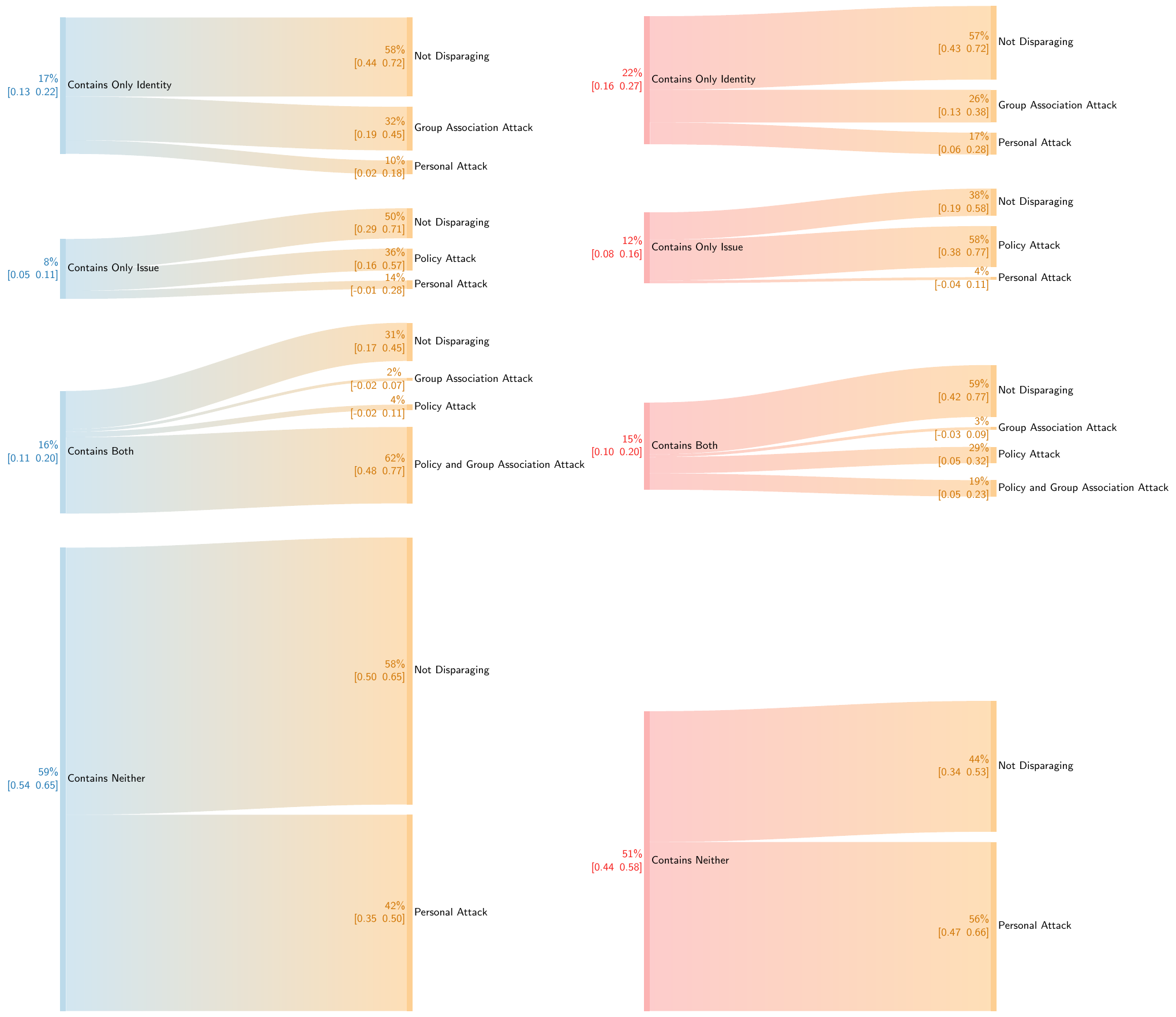}
    \caption{Robustness check: added the 6 realdonaldtrump posts not seen by the sock puppet accounts. Harris (left) and Trump (right) TikTok posts in terms of containing identity, issue, and disparaging content. We see that 50\% of Harris posts and 53\% of Trump posts contain disparaging content. Among all Harris posts, 12\% are policy attacks, 15\% are group association attacks, and 26\% are personal attacks. Among all Trump posts, 12\% are policy attacks, 9\% are group association attacks, and 36\% are personal attacks. The width of nodes represents the number of posts. Standard errors are in parentheses.}
    \label{fig:basics2}
\end{figure}

\begin{figure}[H]
    \centering
    \includegraphics[width=.9\linewidth]{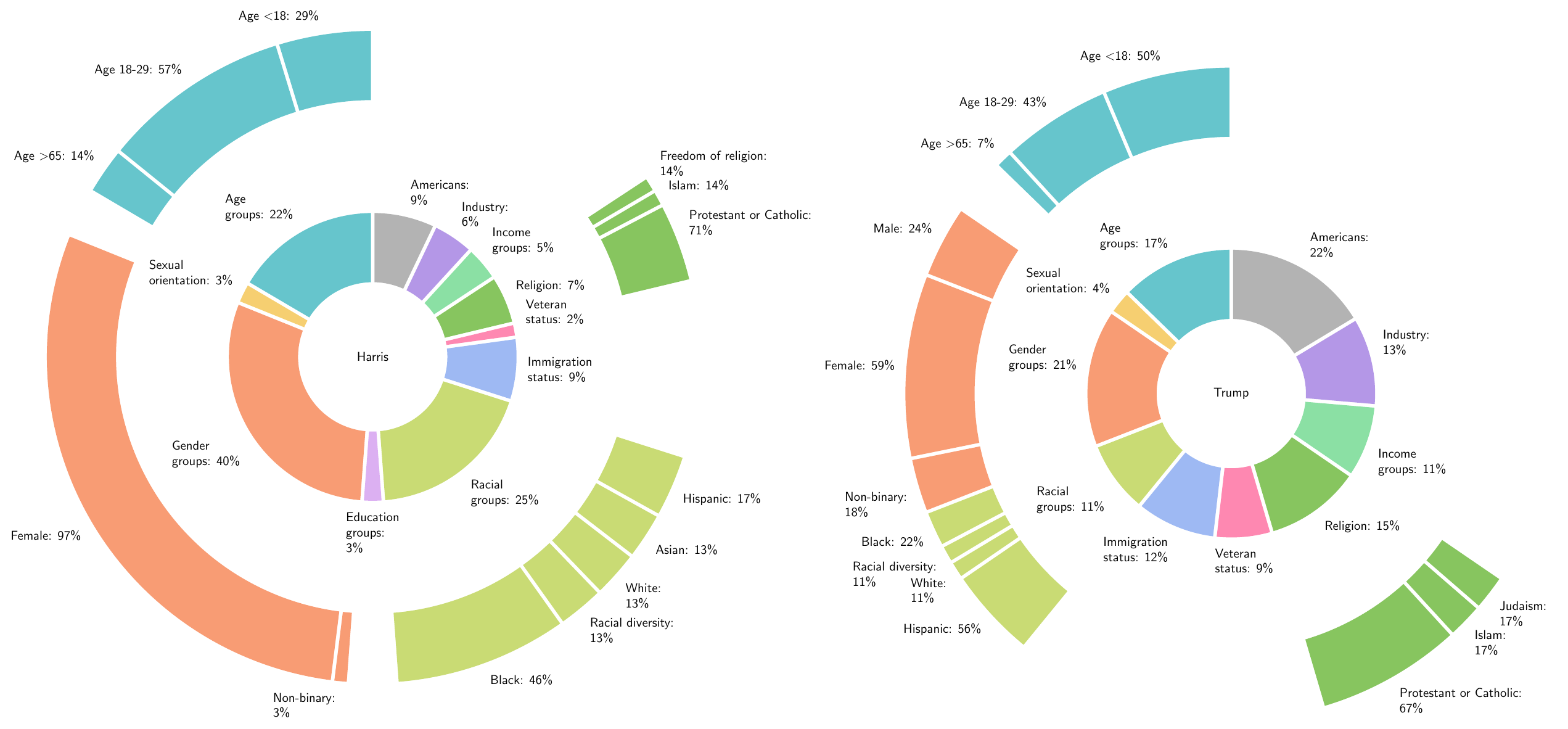}
    \caption{Robustness check: added the 6 realdonaldtrump posts not seen by the sock puppet accounts. The social identities mentioned by Harris and Trump within TikTok posts. Here, we tagged explicit mentions and did not tag a group if it was referred to implicitly (e.g. by the camera panning over members of that group). }
    \label{fig:smids2}
\end{figure}

\begin{figure}[H]
    \centering
    \includegraphics[width=1\linewidth]{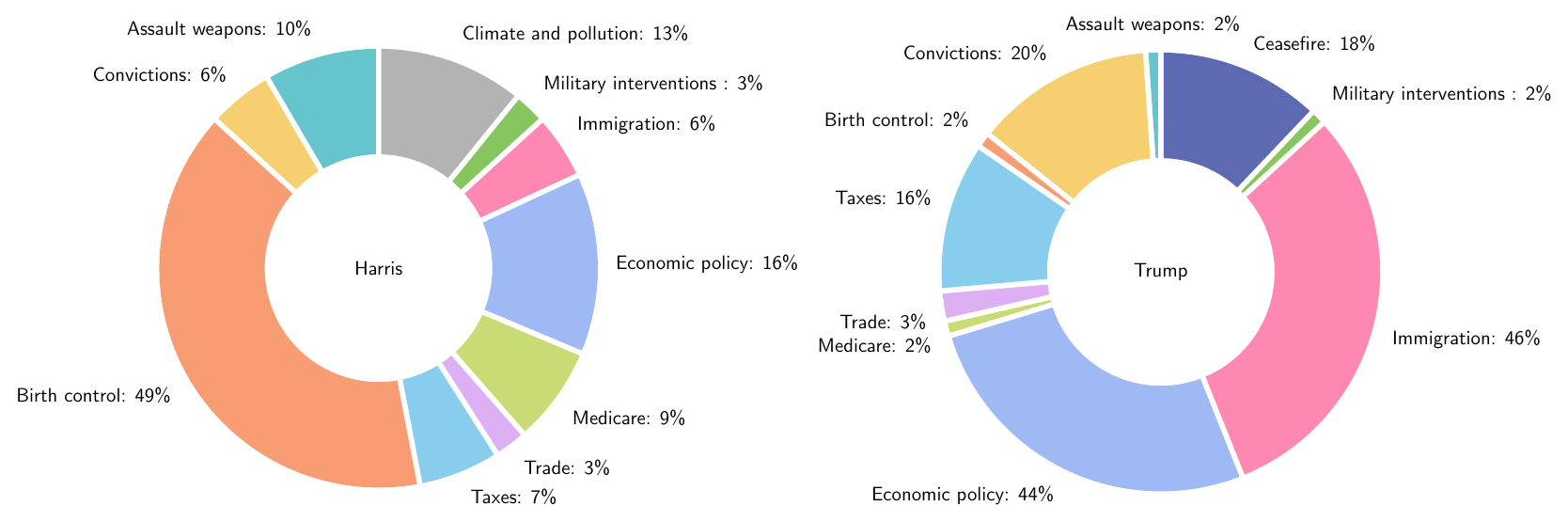}
    \caption{Robustness check: added the 6 realdonaldtrump posts not seen by the sock puppet accounts. The policy issues mentioned by Harris and Trump within TikTok posts. Unlike social identities, here we do not break the issue categories into more granular issues as the candidates do not often discuss these issues in sufficient detail to do so. }
    \label{fig:issues2}
\end{figure}

In this section, we added the 6 realdonaldtrump posts not seen by the sock puppet accounts into analysis and redid Figures \ref{fig:basics}, \ref{fig:smids}, and \ref{fig:issues} as Figures \ref{fig:basics2}, \ref{fig:smids2}, and \ref{fig:issues2}.
We see the qualitative findings are similar to those in the main paper. 

\newpage
\begin{figure}[hb!]
    \centering
    \includegraphics[width=\linewidth]{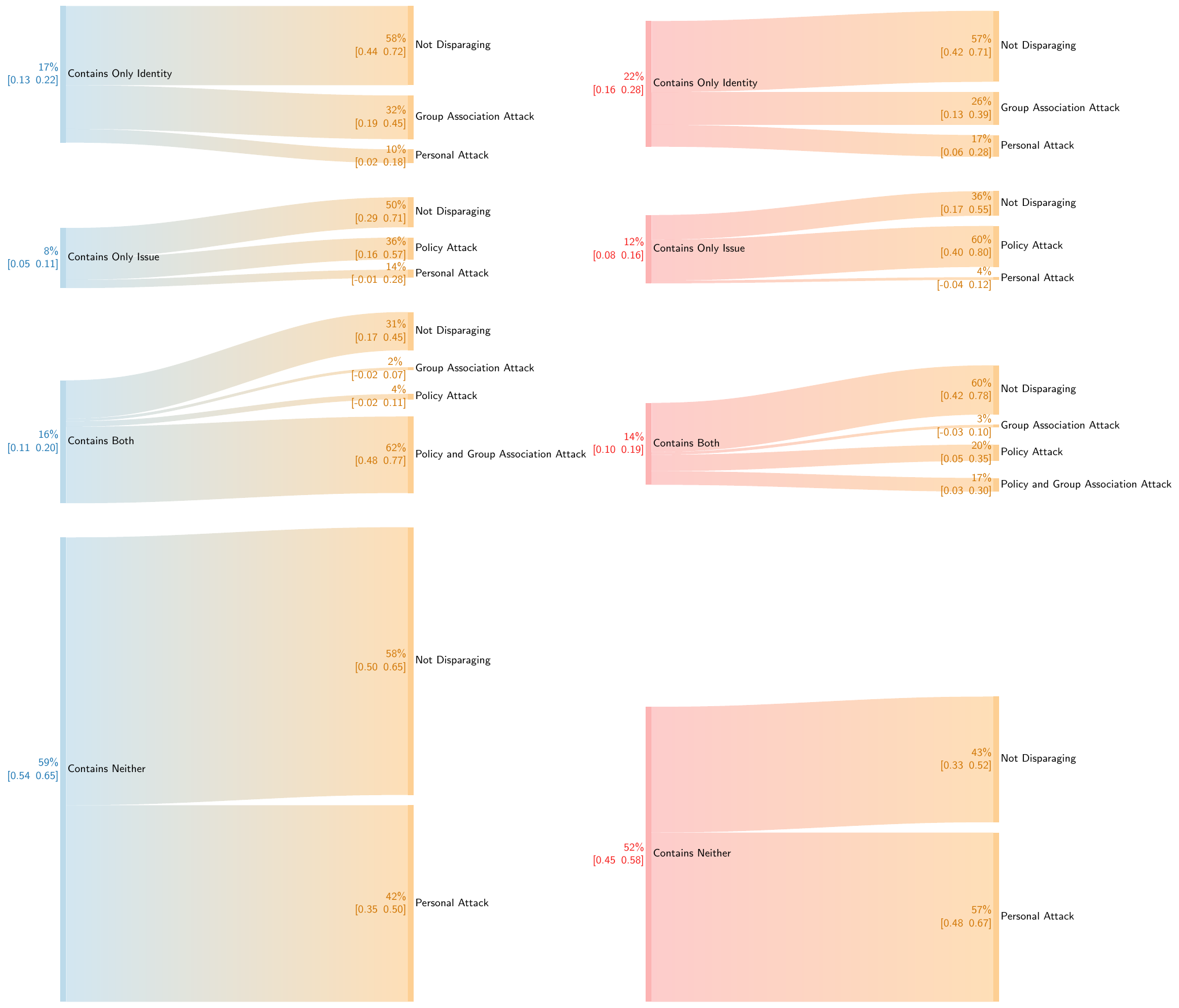}
    \caption{Figure \ref{fig:basics} with 95\% confidence intervals: Harris (left) and Trump (right) TikTok posts in terms of containing identity, issue, and disparaging content. We see that 50\% of Harris posts and 54\% of Trump posts contain disparaging content. Among all Harris posts, 12\% are policy attacks, 15\% are group association attacks, and 26\% are personal attacks. Among all Trump posts, 12\% are policy attacks, 8\% are group association attacks, and 33\% are personal attacks. The width of nodes represents the number of posts. Standard errors are in parentheses.}
    \label{fig:basics3}
\end{figure}

\newpage

\newpage

\subsection{To what extent do social identities drive issue positions and importance?}
\label{identitytoissues}
While not essential to the central questions of the paper we are also interested in understanding whether social identities drive issue positions? 
As we see that the candidates are more likely to discuss identities than issues, it may be that they do so as identities are important in how people form opinions about issues. Thus, we investigate two hypotheses here. 

Here, we are interested in understanding the relationship between social group identities and issue position and importance 
above and beyond partisanship. 
That is, we are teasing apart two hypotheses: 

\begin{itemize}
    \item \textbf{Hypothesis one:} The reason that issues are so predictive is that people take issue positions that align perfectly with the positions of the party.  If this is the case there shouldn't be a correlation between being affected by an issue and one's position about that issue, or the importance they give to it. Instead, we should see that no identity covariates are significant after including party ID in our models of issue position and importance.
    \item \textbf{Hypothesis two:} The reason that issues are so predictive is that people take the party line on issues they don't care about and have more diverse opinions about issues which are important to them. If this is the case, we should see more deviation from the party line for issues that clearly impact one of the social groups that a respondent belongs to. We should see this even when we include party ID in our models of issue position and importance.
\end{itemize}

\begin{figure}
    \begin{center}
    \includegraphics[width=.76\columnwidth]{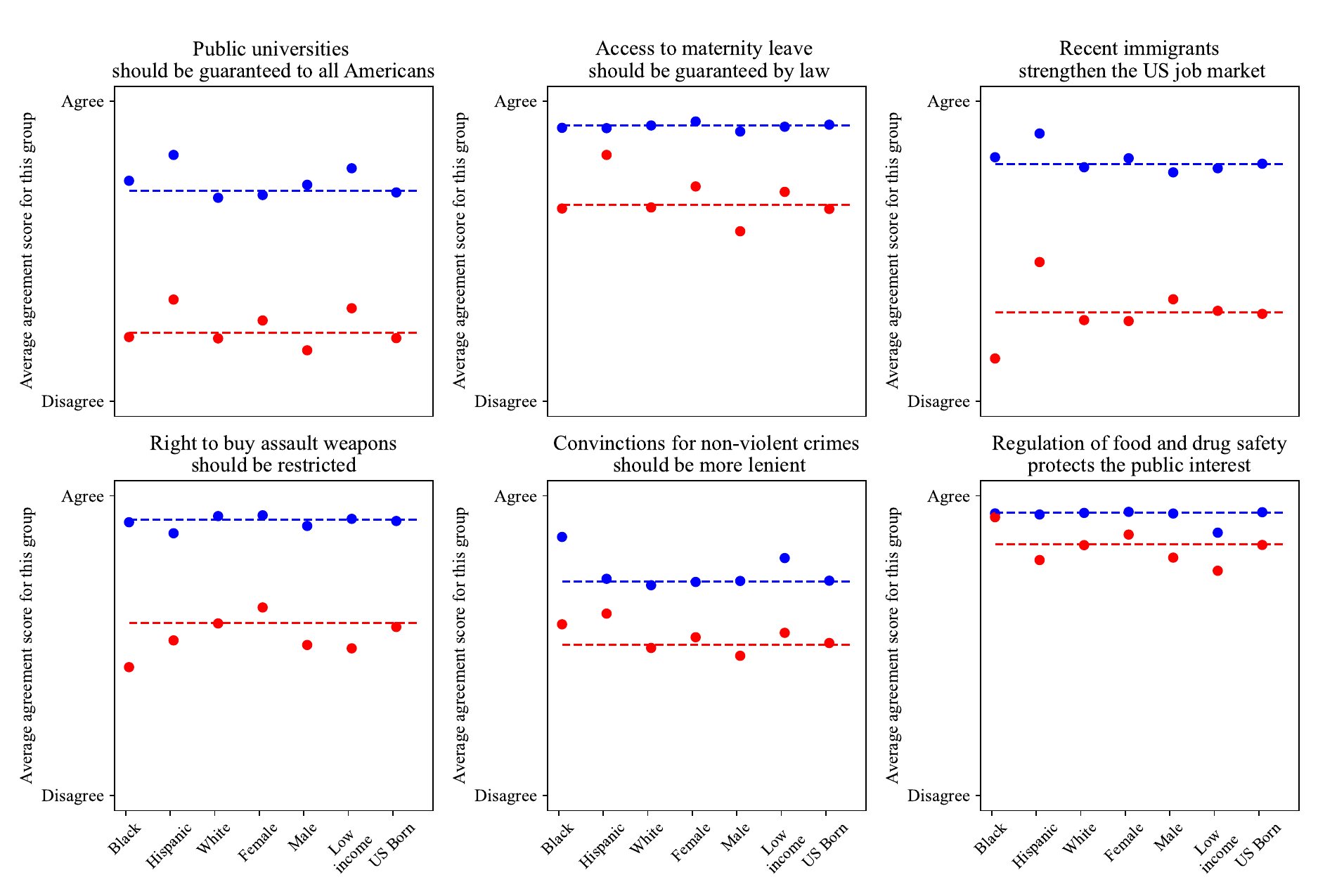}
    \caption{Here we show the average issue position amongst Democrats and Republicans with different demographics. Higher values indicate a more positive stance on this issue.}
    \label{fig:issueposition}
    \end{center}
\end{figure}

In \figref{fig:issueposition}, we see that issue positions seem to be much more determined by party than 
by demographic. However, there are a few notable exceptions. For example, 
Hispanic and low-income respondents have more positive positions towards free education and basic income than the party baseline, within both parties. We see clear party separation on the issues of education, assault weapons, and immigration and less separation on the issues of maternity leave, non-violent crimes, and regulations. 
We see some of these results mirrored in \tabref{tab:positions} where we marked statistical significance of identities to predict issue positions, conditional on party identification. For example, when it comes to the issues of  basic income, we see that Hispanic respondents are more positive on the issue, even when party ID is included in the model. Females are more likely to support maternity leave, when when party ID is considered. Critically, while no identity is significant for all issues, most identity categories are significant for some issues (i.e.,  scanning left to right one can see all categories but sexual orientation are significant for some issue), and some identity categories are significant for all issues (i.e., scanning top to bottom every issue has a least two significant identity categories).

\begin{figure}
    \begin{center}
    \includegraphics[width=.76\columnwidth]{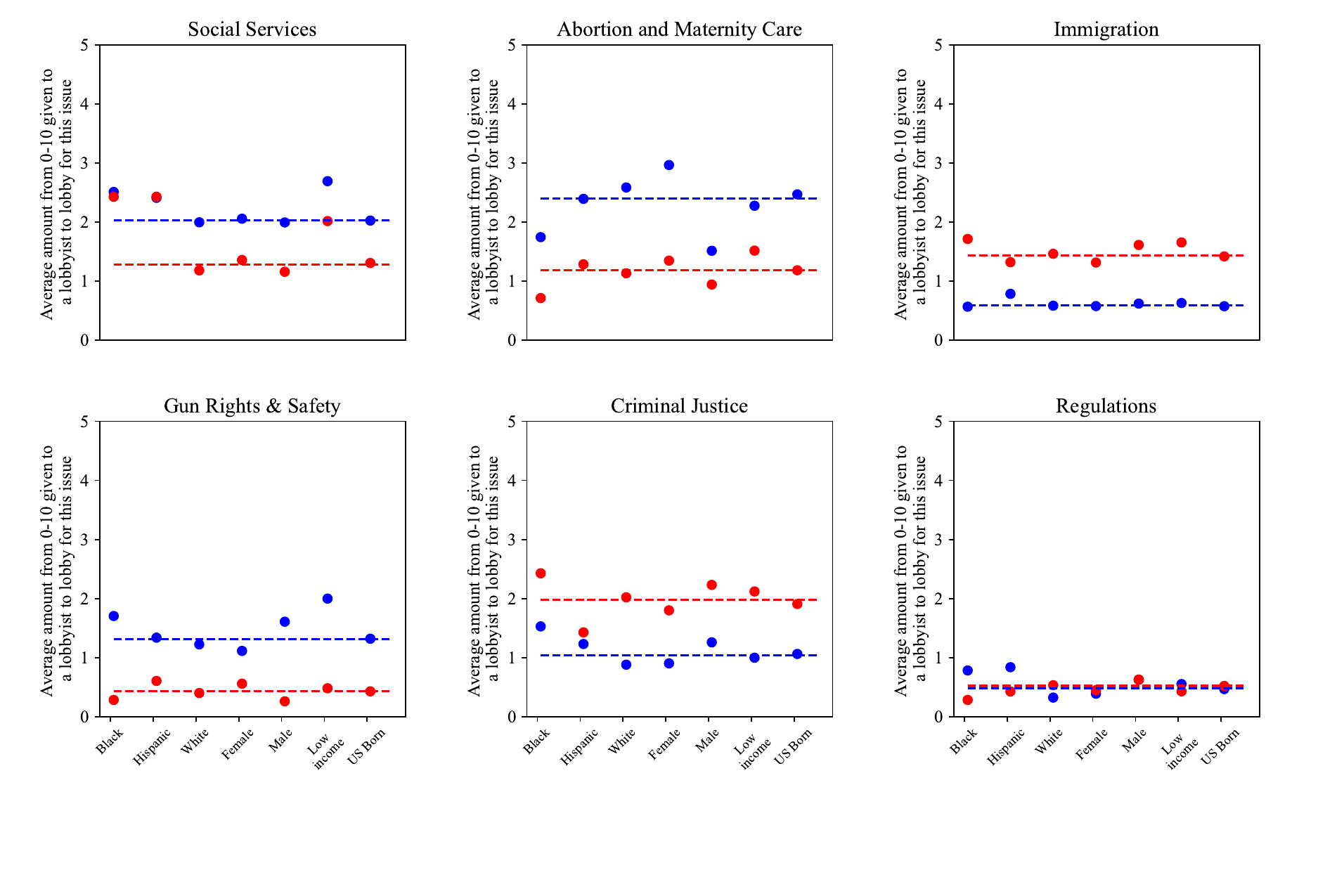}
    \caption{Here we show the average issue area importance amongst Democrats and Republicans with different demographics. Higher values indicate more importance given to a specific area.}
    \label{fig:issueimportance}
    \end{center}
\end{figure}

In \figref{fig:issueimportance}, we see that there are more instances of respondents ``crossing 
the party line''.  For example, we see that Black, Hispanic, and low-income Republicans are more similar to their Democratic counterparts than to the average Republican in terms of the importance they give to the issue area of social services. When it comes to abortion and maternity care Democratic men give similar importance to the issue as Republican men. 
We follow up on whether any of the findings in \figref{fig:issueimportance}
persist when we model issue importance as a function of one's identities \textit{and} their party ID in \tabref{tab:feature_selection}. Here, we see that many of the results are mirrored. Again, females place greater importance on abortion and maternity leave, even when party ID is considered. Low income respondents place greater importance on income inequality and social services. Hispanic respondents also place greater importance on social services. 
%For example, 

\begin{table}[]
    \centering
   \adjustbox{max width=\columnwidth}{%
\begin{tabular}{|l|cccccccc|}
  \cline{1-9}
       & Over the counter    & Free   & Raise  & Pro-immigrants  & Increase cc  & Stricter  &  donations & Non-violent  \\ 

 & birth control    &  education  &  inheritance taxes &  jobs &  spending &  gun laws &   donations & crimes \\ 
       
  \cline{1-9}
      Female&+X&&&&+X&&+X&\\
         % \cdashline{2-9}
         Asian&&&&&&&&\\
      Black&&&&&&&&+X\\
      Hispanic&&+X&&+X&&&&+X\\
    %  White&&&&&&&&\\
         % \cdashline{2-9}
      18-29&&+X&-X&&&&-X&\\
      30-44&&+X&-X&&&&-X&+X\\
      45-64&&+X&&&&-X&&\\
  %    65\&over&&&&&&&&\\
         % \cdashline{2-9}
      HS or less&&&&&&&&\\
  %    Some college&&&&&&&&\\
      4-year degree&&&&&+X&&+X&\\
      Post-grad&&&&+X&&&&\\
         % \cdashline{2-9}
      %Healthcare&&&&&&&&\\
      %Manufacturing&&&&&&&&\\
      %Other Job&&&&+X&&&&\\
      A large city&&&&&&&&\\
    %  A suburb&&&&&&&&\\
     A small town&-X&&&&&&&\\
      A rural area&&&&&&-X&&\\
            % \cdashline{2-9}
      Protestant faith&&&&&&&&\\
      Catholic faith&&&&-X&&&&\\
      Other faith&X&+X&&&&&&\\
                    % \cdashline{2-9}
     % Heterosexual&&&&&&&&\\
      LGBTQ&&&&&&&&\\
                              % % \cdashline{2-9}
      Union member&&&&&&&&-X\\
                              % \cdashline{2-9}
      Army Status&&&-X&&&-X&&\\
                            % \cdashline{2-9}
     US Born&&&-X&&&&&\\
                      % \cdashline{2-9}
     Income $<$ \$25K&&&+X&&&&&+X\\
     Income $\in$ \$25-50K&&&+X&&&&&+X\\
     Income $\in$ \$50-90K&&&&&&&&\\
     Income $\in$ \$90-250K&&&&+X&&&&\\
     Income $>$ \$250K&+X&&&+X&&&&\\
                        % \cdashline{2-9}
     Independent&-X&-X&-X&-X&-X&&&-X\\
     Republican&-X&-X&-X&-X&-X&-X&-X&-X\\
  \cline{1-9}
    \end{tabular}
}
    \caption{We fit a separate linear regression for each issue using all identity features and party ID as covariates. We find that for each issue some covariates are statistically significant when when we include party ID as a coefficient. Here an X indicates that a coefficient was significant, a + indicates a positive coefficient and a - indicates a negative coefficient.}
    \label{tab:positions} 
\end{table}

\begin{table}[]
\centering
    \adjustbox{width=\columnwidth}{%
    
\begin{tabular}{  |l|*{8}{c} |}
  \cline{1-9}
       &Abortion   & Social Services & Immi-gration  & Criminal Justice & Income Inequality & Voting Rights & Taxes & Environment  \\ 
  \cline{1-9}
      Female&+X&&&-X&&&&\\
                     % \cdashline{2-9}
      Asian&&&&&&&&\\
      Black&&&&&&&&-X\\
      Hispanic&&+X&&&&&-X&\\
    %  White&&&&&&&&\\
                   % \cdashline{2-9}
      18-29&&&&&+X&&&\\
      30-44&+X&&-X&&+X&-X&&\\
      45-64&&&&&+X&&&\\
  %    65\&over&&&&&&&&\\
                 % \cdashline{2-9}
      HS or less&&+X&&-X&&&&\\
  %    Some college&&&&&&&&\\
      4-year degree&&&&&&&&\\
      Post-grad&&&&&&&&\\
                     % \cdashline{2-9}
      %Healthcare&&&&&&&&\\
      %Manufacturing&&&&&&&&\\
      %Other Job&&&&+X&&&&\\
      A large city&&&&&&&&\\
    %  A suburb&&&&&&&&\\
    A small city&&&&&&&&\\
      A rural area&&&&&&&&\\
                     % \cdashline{2-9}
      Protestant faith&&&&&&&&\\
      Catholic faith&&&&&&&+X&\\
      Other faith&+X&&&&&&&\\
                     % \cdashline{2-9}
     % Heterosexual&&&&&&&&\\
      LGBTQ&&&&&+X&&&\\
                     % \cdashline{2-9}
      Union member&&+X&&&&&&\\
                     % \cdashline{2-9}
      Army Status&&&&&&&&\\
                     % \cdashline{2-9}
     US Born&&&&&&&+X&\\
                    % \cdashline{2-9}
     Income $<$ \$25K&&+X&&&+X&&&\\
     Income $\in$ \$25-50K&&&&&+X&&&\\
     Income $\in$ \$50-90K&&&&&&&&\\
     Income $\in$ \$90-250K&&&&&&&&\\
     Income $>$ \$250K&&&&&&&&\\
                    % \cdashline{2-9}
     Independent&&&+X&+X&&-X&+X&-X\\
     Republican&-X&-X&+X&+X&-X&&+X&-X\\
  \cline{1-9}
    \end{tabular}
    }
    \caption{We fit a separate linear regression for the amount of fictional lobbying credits placed on each issue using all identity features and party ID as covariates. We find that for most issue importances some covariates are statistically significant when when we include party ID as a coefficient. Here an X indicates that a coefficient was significant, a + indicates a positive coefficient and a - indicates a negative coefficient.}
    \label{tab:feature_selection}
\end{table}

Generally, these results suggest greater support for hypothesis two. Party identification is undeniably correlated with both issue 
positions and importance. However, we do see some deviations from the party line, especially when 
it comes to how respondents value different issues. Although it is not our central question, we lend some evidence towards the claim that policy influences party rather than the other way around, 
by showing that how one values different policies has a relationship to the groups one belongs to 
and is not strictly determined by one's party. 

Yet, we caution to attribute identity to the findings shown here. We found almost no correlation between identity strength and issue importance. Here, we measure identity strength as whether you respond that belonging to a certain group is important to your identity or if you see the fate of that group as linked 
to your fate. 
Across the board, identity strength did not impact policy position or importance, with a few exceptions. 
Only in two policy areas is identity centrality 
more correlated to policy area preferences
than simply belonging to an affected group:
reproductive rights, and immigration.
We find that women who view gender as more central to their
identity are more likely to prioritize the policy area 
of reproductive rights. Additionally, 
people born in the US who place being American 
as central to their identity are 
more likely to prioritize immigration as a policy area.
However, these are the only statistically significant 
relationships that we find. Generally, 
we find that certain issue positions are slightly affected
by group belonging, and largely unaffected
by identity strength.
\end{appendices}

\end{document}